\documentclass[a4paper,11pt]{article}

\usepackage[utf8]{inputenc}
\usepackage[T1]{fontenc}

\usepackage{jcapmod}

\usepackage{amsmath,amsfonts,amssymb}
\usepackage{physics}
\usepackage{mathtools}
\usepackage{bm}
\usepackage{cancel}

\usepackage{hyperref}

\usepackage{graphicx}

\usepackage[a4paper,margin=2.5cm]{geometry}

\numberwithin{equation}{section}

\begin{document}

\pagenumbering{roman}

\begin{titlepage}

\baselineskip=15.5pt \thispagestyle{empty}

\begin{center}
    {\fontsize{19}{24}\selectfont \bfseries Cosmology at the top of the $\alpha'$ tower}
\end{center}

\vspace{0.1cm}

\begin{center}
    {\fontsize{12}{18}\selectfont Jerome Quintin,$^{1}$ Heliudson Bernardo,$^{2}$ and Guilherme Franzmann$^{3}$}
\end{center}

\begin{center}
    \vskip8pt
    \textsl{$^1$ Max Planck Institute for Gravitational Physics (Albert Einstein Institute),\\
        Am M\"uhlenberg 1, D-14476 Potsdam, Germany}
    \vskip8pt
    \textsl{$^2$ Department of Physics, Ernest Rutherford Physics Building, McGill University,\\
        3600 rue Universit\'e, Montr\'eal, Qu\'ebec H3A 2T8, Canada}
    \vskip8pt
    \textsl{$^3$ Nordita, KTH Royal Institute of Technology and Stockholm University,\\
Hannes Alfvéns v\"ag 12, SE-106 91 Stockholm, Sweden}
\end{center}

\vspace{1.2cm}

\hrule
\vspace{0.3cm}

\noindent {\bf Abstract}\\[0.1cm]
The cosmology of the fully $\alpha'$-corrected duality-invariant action for the Neveu-Schwarz sector of string theory is revisited, with special emphasis on its coupling to matter sources. The role of the duality covariant pressure and dilatonic charge of the matter sector is explored in various contexts, from the low-curvature regime to non-perturbative solutions in $\alpha'$. We comment on how an infinite tower of $\alpha'$ corrections allows for fixed-dilaton de Sitter solutions, even in vacuum. We further investigate the necessary conditions for accelerated expansion in the Einstein frame, as well as for non-singular bounces that could resolve the big bang singularity. In particular, explicit examples are constructed, which show that the tower of $\alpha'$ corrections may support an Einstein-frame non-singular cosmological bouncing background, even when the matter sector respects the null energy condition.
\vskip10pt
\hrule
\vskip10pt

\end{titlepage}

\thispagestyle{empty}

\setcounter{page}{2}
\tableofcontents
\newpage
\pagenumbering{arabic}
\setcounter{page}{1}
\clearpage



\section{Introduction}

Despite the enormous theoretical and observational success of Einstein's general theory of relativity, one major limitation is its failure to describe physics at very high-curvature scales. Classically, the theory predicts that, under common assumptions, spacetime manifolds inevitably possess singular points where measurable quantities unphysically diverge \cite{Penrose:1964wq,Hawking:1967ju,Hawking:1969sw}. Quantum mechanically, general relativity (GR) is non-renormalizable: diverging graviton loops always require the addition of higher-curvature counterterms in the action (up to arbitrarily high powers) \cite{tHooft:1974toh,Christensen:1979iy,Goroff:1985th}.

One of the hopes of string theory is precisely to resolve these shortcomings. By promoting point particles living at singular points in spacetime to extended objects like strings, one can imagine that divergences and singularities should be resolved. Given the extended nature of strings, scattering amplitudes described by one-dimensional graphs in quantum field theory (namely Feynman diagrams) are replaced by two-dimensional manifolds (namely string worldsheets),
thus removing the sharp localization of interaction vertices. Furthermore, the only dimensionful parameter of the theory can be associated with the fundamental length of a string, $\ell_\mathrm{s}$ (often equivalently characterized by the $\alpha'$ parameter via $\ell_\mathrm{s}\equiv\sqrt{2\pi\alpha'}$), thus suggesting the existence of a minimal length scale -- a recurring property of attempts to finding a quantum theory of gravity \cite{Hossenfelder:2012jw} -- that would indicate the boundedness of spacetime curvature.
However, it remains to be shown that singular spacetimes can be fully and consistently resolved in a fundamental theory. 
For instance, we do not know yet if string theory fundamentally allows for non-singular bouncing cosmologies that resolve the big bang.

String theory is most easily approached as a perturbative theory in $\alpha'$ as it is the only dimensionful parameter at play. To lowest order in $\sqrt{\alpha'}/\mathcal{R}\ll 1$, where $\mathcal{R}$ is the characteristic curvature radius of the background,
consistency of the sigma model by ensuring the vanishing of the Weyl anomaly (correspondingly enforcing the one-loop beta function to vanish) yields, in a purely gravitational theory, $\beta_{\mu\nu}=\alpha' R_{\mu\nu}=0$, which is precisely the vacuum Einstein equation. Going to next order in $\sqrt{\alpha'}/\mathcal{R}$, one can compute the two-loop beta function yielding $\beta_{\mu\nu}=\alpha' R_{\mu\nu}+(\alpha^{\prime\,2}/2)R_{\mu\gamma\rho\sigma}R_\nu{}^{\gamma\rho\sigma}$. It thus becomes apparent how $\alpha'$ corrections introduce higher-curvature terms in the equations of motion (and equivalently in the action). While the calculation may be tackled up to a few loops (and in principle to all orders) \cite{Gross:1986iv,Gross:1986mw,Metsaev:1987zx}, subtleties rapidly arise and calculations become very difficult. To make matters worse, the demonstration that singularities are truly removed in the theory might only be achieved from a non-perturbative theory that encapsulates $\alpha'$ corrections to all orders, i.e.~a theory with an infinite number of higher-curvature terms that can probe the regime of the fundamental scale of the theory, $R\sim\ell_\mathrm{s}^{-2}$, which is precisely the non-perturbative regime of the theory.

Fortunately, string theory comes with useful features such as dualities. In particular, T-duality implies the correspondence between physics at scale $\mathcal{R}$ with that at scale $\alpha'/\mathcal{R}$. This is manifest for instance in a cosmological setup (when one imposes the symmetries that come with the assumptions of homogeneity and isotropy) described by a scale factor $a(t)$: the physics of the universe with scale factor $a$ is dual to that with scale factor $1/a$. This is known as the discrete scale factor duality \cite{Veneziano:1991ek}, which is generalized to the continuous group\footnote{Here and throughout, we assume the group with real elements, i.e., $\mathrm{O}(d,d)$ is meant to be shorthand notation for $\mathrm{O}(d,d,\mathbb{R})$.} $\mathrm{O}(d,d)$ in $d$ spatial dimensions \cite{Meissner:1991zj,Tseytlin:1991wr}. The duality is proved to hold
to all order in $\alpha'$ \cite{Sen:1991zi} (see \cite{Bergshoeff:1995cg,Meissner:1996sa,Elgood:2020xwu,Codina:2020kvj} for explicit verifications at finite orders in $\alpha'$), hence suggesting $\mathrm{O}(d,d)$ symmetry may be imposed at the action level to help one construct a cosmological theory\footnote{An alternative approach to finding $\alpha'$ corrections by taking advantage of T-duality has been developed thanks to double field theory \cite{Hohm:2013jaa,Hohm:2014eba,Hohm:2014xsa,Marques:2015vua,Hohm:2016lim,Lescano:2016grn,Hohm:2016lge,Hohm:2016yvc,Baron:2017dvb,Baron:2018lve,Baron:2020xel,Hronek:2020xxi} (see also \cite{Angus:2018mep,Park:2019hbc}). This benefits from being a spacetime-covariant approach, which does not have to rely on the assumptions of homogeneity and isotropy. This avenue has led to promising applications in cosmology \cite{Wu:2013sha,Brandenberger:2018xwl,Brandenberger:2018bdc,Bernardo:2019pnq,Angus:2019bqs}. While this path will not be explored in the present work, we believe that it is worthy of mention. In passing, it is also worth mentioning additional approaches to constructing $\alpha'$-corrected stringy actions and their application to stringy black holes (see, e.g., \cite{Ortin:2020xdm,Cano:2017qrq,Cano:2018qev,Cano:2019ycn}).} that could include $\alpha'$ corrections to all orders. This is what was achieved by Hohm \& Zwiebach \cite{Hohm:2019jgu,Hohm:2019ccp} (generalizing \cite{Hohm:2015doa}), who found the form of the most general action for the metric,
the dilaton and the $b$-field when these fields only depend on time, which is invariant under $\mathrm{O}(d,d)$ transformations and which includes $\alpha'$ corrections to all orders. The inclusion of coupling to other matter fields was then first explored in \cite{Bernardo:2019bkz}, and applications to string cosmology have been explored in a few studies so far \cite{Hohm:2019jgu,Hohm:2019ccp,Krishnan:2019mkv,Bernardo:2019bkz,Bernardo:2020zlc,Bernardo:2020nol,Bernardo:2020bpa,Nunez:2020hxx}, often discovering new potential solutions, such as de Sitter-like or loitering backgrounds, which are otherwise often difficult to find in string cosmology (see, e.g., \cite{Obied:2018sgi,Agrawal:2018own,Battefeld:2005av,Brandenberger:2008nx}).

The current paper's aim is to go beyond these recent works on string-cosmology backgrounds to all orders in $\alpha'$. The precise goal shall be twofold: putting recent developments on firmer grounds and addressing the question of cosmological singularity resolution. The former is relevant since it has been realized that subtleties arise when coupling matter to an $\mathrm{O}(d,d)$-invariant theory with $\alpha'$ corrections to all orders, and this plays an important role when it comes to describing physical cosmological scenarios with matter. In particular, we unveil that pressure can have different definitions, and we show the role of symmetries in constraining allowed theories. From this, we give a proof that if the theory does not couple to external fields (or even with matter satisfying some specific equation of state [EoS]), constant-dilaton solutions only admit Minkowski or de Sitter (dS) backgrounds. As a consistency test, we also show how the linear dilaton conformal field theory (CFT) is recovered in the present theory. Furthermore, we explore the role of matter in the different regimes of the theory; specifically, we show how the low- and high-energy limits can fix the allowed equations of state of matter. The latter is most easily tackled from the point of view of energy conditions, which are at the core of the singularity theorems in GR \cite{Penrose:1964wq,Hawking:1967ju,Hawking:1969sw}. Indeed, given some matter content, one can derive effective energy conditions such as the strong energy condition (SEC) or the null energy condition (NEC), and the violation of these conditions are often necessary for finding certain cosmological solutions, such as accelerating and non-singular ones. From this, we can comment on the viability of the theory to yield dS-like accelerating solutions  and finally address the question of singularity resolution. Specifically, we explore different physically motivated ans\"{a}tze and the resulting phase space of the theory, seeking for consistent non-singular solutions. The results shall be interestingly nuanced: previously claimed non-singular cosmological solutions are actually ruled out, but we find other ans\"{a}tze that do admit, e.g., non-singular cosmological bounces; however, fully singularity-free solutions remain hard to find and are at the level of toy models, and so, it may indicate that singularity resolution might not be so straightforwardly implied from string theory, even with $\alpha'$ corrections to all orders.

\paragraph*{Outline}
We start by revisiting $\alpha'$-complete cosmology in Sec.~\ref{sec:revisit}, first reviewing the basics of the theory and the resulting equations of motion in vacuum (Sec.~\ref{sec:review}) and with the addition of matter (Sec.~\ref{sec:matter_sector}). Then, Sec.~\ref{sec:linktononOdd} clarifies the different definitions of pressure in an $\mathrm{O}(d,d)$-invariant theory. The low- and high-curvature limits of the theory are discussed in Sec.~\ref{sec:lims}, and the linear dilaton CFT is presented in Sec.~\ref{sec:linear_dilaton}. Constant-dilaton solutions and their implications for dS spacetime are carefully studied in Sec.~\ref{sec:constantdilaton}. The topic of energy conditions is the subject of Sec.~\ref{sec:SEC_disc}, exploring both the SEC (Sec.~\ref{sec:SEC}) and NEC (Sec.~\ref{sec:NEC}), their violation and the corresponding implications. Section \ref{sec:bouncing} is devoted to finding non-singular bouncing cosmological backgrounds, deriving general requirements (Sec.~\ref{sec:bouncegen}) and exploring specific models (Sec.~\ref{sec:ansaetze}). We summarize the results in Sec.~\ref{sec:discussion}, together with further discussions.

\paragraph*{Notation} Throughout this paper, we follow as much as possible the notation of \cite{Hohm:2019jgu,Bernardo:2019bkz}, though with some slight differences.
In particular, let us use sans-serif fonts to denote general matrices (e.g., $\mathsf{G}$), bold italic for spacetime tensors of any rank (e.g., $\bm{G}$), and standard italic for scalar quantities (in the mathematical sense) and tensorial components with indices (e.g., $n$, $g_{ij}$, $G_{\mu\nu}$).
Also, there are $D=d+1$ spacetime dimensions, $\mu,\nu,\ldots\in\{0,1,\ldots,d\}$, $i,j,\ldots\in\{1,\ldots,d\}$, and we use the mostly plus signature $(-,+,\ldots,+)$.

\section{Revisiting $\alpha'$-complete cosmology and coupling to matter}\label{sec:revisit}

\subsection{Brief review of the gravitational $\mathrm{O}(d,d)$-invariant action}\label{sec:review}

The universal massless Neveu-Schwarz sector of all superstring theories is composed of three fields: the symmetric spacetime metric tensor ($\bm{G}$), the antisymmetric Kalb-Ramond tensorial field ($\bm{B}_{(2)}$), and the dilaton scalar field ($\phi$). Upon considering a purely time-dependent $(d+1)$-dimensional string background,
\begin{equation}
    G_{\mu\nu}\mathrm{d}x^\mu\mathrm{d}x^\nu=-n(t)^2\mathrm{d}t^2+g_{ij}(t)\mathrm{d}x^i\mathrm{d}x^j\,, \quad \bm{B}_{(2)} = b_{ij}(t) \mathbf{d}x^i\wedge \mathbf{d}x^j\,, \quad \phi = \phi(t) \,,\label{eq:back}
\end{equation}
these fields can be reassembled to form field representations of the $\mathrm{O}(d,d)$ symmetry group,
\begin{subequations}
\begin{align}
      \Phi &= 2\phi-\ln\sqrt{g}\,, \label{eq:shifteddilatondef} \\
      \mathsf{S} &=\begin{pmatrix}
                \mathsf{b g^{-1}} & \mathsf{g - bg^{-1}b} \\
                \mathsf{g}^{-1} & \mathsf{-g^{-1}b}\end{pmatrix} \,, \label{eq:Smatrixdef}
\end{align}
\end{subequations}
where the shifted dilaton $\Phi$ is an $\mathrm{O}(d,d)$-invariant scalar field, and $\mathsf{S}$ is an $\mathrm{O}(d,d)$-covariant $2d\times 2d$ matrix constructed out of the matrix components of the two-form field $b_{ij}$, the spatial metric $g_{ij}$ and its inverse $g^{ij}$. Note that $g:=\mathrm{det}(\mathsf{g})$ and $\sqrt{-G}=n\sqrt{g}$, with $G:=\mathrm{det}(\mathsf{G})$. The lapse function is denoted by $n(t)$.

Starting with the background ansatz \eqref{eq:back}, further assuming isotropy and flatness (so $g_{ij}(t)=a(t)^2\delta_{ij}$, i.e.~taking a Friedmann-Lema\^itre-Robertson-Walker [FLRW] metric)
and setting the two-form field to zero (so $b_{ij}(t)\equiv 0$),
the $\mathrm{O}(d,d)$-invariant action for these fields takes the general form \cite{Hohm:2019jgu}
\begin{equation}
    S[\Phi,n,\mathsf{S}]=\frac{1}{2\kappa^2}\int\mathrm{d}t\,ne^{-\Phi}\left(-(\mathcal{D}_t\Phi)^2+\sum_{k=1}^\infty(\alpha')^{k-1}c_k\mathrm{tr}\left[(\mathcal{D}_t\mathsf{S})^{2k}\right]\right)\,, \label{eq:action_vacuum}
\end{equation}
where a spatial-volume integral is omitted since everything only depends on time, $\kappa^2 \propto \ell_\mathrm{s}^{d-1}$ essentially defines the $D$-dimensional Newton constant (the proportionality factor varies according to the 10-dimensional theory considered), $\mathcal{D}_t:= n^{-1}\partial_t$ defines the time-reparametrization invariant time derivative, $c_1 = -1/8$ according to the known tree-level, low-energy effective action of bosonic string theory, and the other $c_k$'s are generally unknown (they depend again on which type of string theory is considered). Extremizing this action with respect to (w.r.t.)~$\Phi$, $n$, and $\mathsf{S}$ yields the equations of motion (EOMs)
\begin{subequations}\label{eq:EOM_vacuum}
    \begin{align}\label{eq:EOM_Phi_vacuum}
         2\mathcal{D}_t^2{\Phi} - (\mathcal{D}_t\Phi)^2 - \sum_{k=1}^{\infty}(\alpha')^{k-1}c_k \mathrm{tr}\left[(\mathcal{D}_t\mathsf{S})^{2k}\right] & =0\,,\\ \label{eq:EOM_n_vacuum}
        (\mathcal{D}_t\Phi)^2 - \sum_{k=1}^{\infty}(\alpha')^{k-1}(2k-1)c_k \mathrm{tr}\left[(\mathcal{D}_t\mathsf{S})^{2k}\right] & =0\,,\\ \label{eq:EOM_S_vacuum}
         \mathcal{D}_t\left(e^{-\Phi}\sum_{k=1}^{\infty}(\alpha')^{k-1}4kc_k\mathsf{S}(\mathcal{D}_t\mathsf{S})^{2k-1}\right) &= 0\,,
    \end{align}
\end{subequations}
respectively. Note that the action \eqref{eq:action_vacuum} takes into account an infinite tower of $\alpha'$ string corrections in a flat cosmological background in the absence of the antisymmetric field.
Although these corrections appear as higher-order derivative operators multiplied by generally unknown coefficients, their form is constrained by the ansatz and the $\mathrm{O}(d,d)$ invariance such that they depend only on the first time derivative of $\mathsf{S}$ (higher-derivative terms can always be recast into single-derivative terms via field redefinitions as shown by \cite{Hohm:2019jgu}). However, imposing a vanishing two-form field has its costs. As one can easily check, the matrix $\mathsf{S}$ transforms under $\mathrm{O}(d,d)$ transformations as a rank-two tensor. Thus, a continuous $\mathrm{O}(d,d)$ rotation mixes $\mathsf{g}$ and $\mathsf{b}$ together. By considering  $\mathsf{b}=\mathsf{0}$, the group is explicitly broken down to the discrete scale factor duality \cite{Veneziano:1991ek,Gasperini:1991ak}. Had the two-form field been non-vanishing, other corrections would have been present as multi-trace operators. The explicit derivation of the EOMs and search for solutions in this case has been recently obtained in \cite{Nunez:2020hxx} in the absence of matter, while another approach which includes matter and specific conditions so that these multi-trace corrections can be ignored has recently been developed in \cite{Bernardo:2021xtr}. Nonetheless, some of the following results apply even with full $\mathrm{O}(d,d)$ symmetry, and these cases will be explicitly discussed.

\subsection{The matter sector and the equations of motion}\label{sec:matter_sector}

The gravitational action \eqref{eq:action_vacuum} coupled to matter in an $\mathrm{O}(d,d)$-invariant fashion is \cite{Bernardo:2019bkz}
\begin{equation}
    S[\Phi,n,\mathsf{S},\chi]=\frac{1}{2\kappa^2}\int\mathrm{d}t\,ne^{-\Phi}\left(-(\mathcal{D}_t\Phi)^2+\sum_{k=1}^\infty(\alpha')^{k-1}c_k\mathrm{tr}\left[(\mathcal{D}_t\mathsf{S})^{2k}\right]\right)+S_\mathrm{m}[\Phi,n,\mathsf{S},\chi]\,,\label{eq:actionfull}
\end{equation}
where $\chi$ represents any additional matter fields (denoted as a scalar here, but there could be many such fields and they
could just as well have different spins).
The matter action, $S_\mathrm{m}$, is assumed to be $\mathrm{O}(d,d)$ invariant.
Then, one can define an energy-momentum tensor of the form
\begin{equation}
    T_{\mu\nu}:=-\frac{2}{\sqrt{-G}}\frac{\delta S_\mathrm{m}}{\delta G^{\mu\nu}} \,. \label{eq:energy_momentum_tensor}
\end{equation}
At this point, since $\Phi$, $n$, and $\mathsf{S}$ are independent fields in an $\mathrm{O}(d,d)$-covariant formalism, the variation w.r.t.~the inverse metric tensor components $G^{\mu\nu}$ is taken keeping the other fields fixed.
Specifically, when evaluating $T_{00}$, one varies w.r.t.~$G^{00}=-n^{-2}$, so one is really varying w.r.t.~$n$. 
Similarly, when evaluating $T_{ij}$, one varies w.r.t.~$G^{ij}=g^{ij}$, so one is really varying w.r.t.~$\mathsf{S}$ [recall Eq.~\eqref{eq:Smatrixdef}]. In doing so, one must keep $n$ and $\Phi$ fixed.
This is counter-intuitive since $\Phi=2\phi-\ln\sqrt{g}$ appears to have explicit $\mathsf{g}$ dependence, but this dependence must be `ignored' to be consistent with how the EOMs are derived in the $\mathrm{O}(d,d)$-invariant formalism. In \cite{Bernardo:2019bkz}, the variations were done w.r.t.~the $\mathrm{O}(d,d)$ multiplet fields, and as a result the variation w.r.t.~the spatial metric at constant $\Phi$ was naturally encoded in a $2d\times 2d$ $\mathrm{O}(d,d)$ tensor, which was a source to $\mathsf{S}$. As is discussed below in Sec.~\ref{sec:linktononOdd}, this is different than what is done in standard [not $\mathrm{O}(d,d)$] string cosmology, nevertheless consistent.\footnote{Indeed, the fact that one is considering $\mathrm{O}(d,d)$ covariance as opposed to diffeomorphism covariance as a starting point leads one to consider field redefinitions that mix
the metric, the dilaton, and the two-form (in general). This is simply a choice, and it is very much similar to what one does in multi-field inflation to distinguish between adiabatic and entropic perturbations or just in cosmological linear perturbation theory, where the physical variable is a gauge-invariant combination of the metric and the matter perturbations.}

One can also define the $\Phi$ charge density by
\begin{equation}
    \sigma:=-\frac{2}{\sqrt{-G}}\left(\frac{\delta S_\mathrm{m}}{\delta\Phi}\right)_\mathsf{S}
    \,, 
    \label{eq:dilaton_density}
\end{equation}
where the same logic as above is considered: the variation w.r.t.~$\Phi$ must be performed keeping the other variables fixed, in particular $n$ and $\mathsf{S}$, even though both $\Phi$ and $\mathsf{S}$ can be thought of as functions of $g_{ij}$.
The purpose of the notation sometimes used as in Eq.~\eqref{eq:dilaton_density} is to make this explicit.

From the definition of the energy-momentum tensor and assuming a perfect fluid, where $T^\mu{}_\nu=\mathrm{diag}(-\rho,p\,\delta^i{}_j)$,
the energy density and the pressure are readily defined by Eq.~\eqref{eq:energy_momentum_tensor} to be
\begin{subequations}
\begin{align}
    \rho & = -\frac{1}{\sqrt{g}} \frac{\delta S_\mathrm{m}}{\delta n} \,, \label{eq:energy_density}\\
    p & = -\frac{2}{n\sqrt{g}} \frac{g^{ij}}{d} 
   \left(\frac{\delta S_\mathrm{m}}{\delta g^{ij}}\right)_\Phi \,. \label{eq:pressure}
\end{align}
\end{subequations}
Finally, considering these matter sources defined in a flat FLRW background (taking $n(t)=1$ and $\mathsf{g}=a(t)^2\mathsf{I}$, where $\mathsf{I}$ is the identity matrix), and using the left-hand side (LHS) of \eqref{eq:EOM_vacuum}, the EOMs reduce to \cite{Bernardo:2019bkz}
\begin{subequations}\label{eq:EOM_matter_final}
    \begin{align}
    \dot{\Phi}^2 + HF'(H) - F(H) & = 2\kappa^2 e^{\Phi}\Bar{\rho}  \label{eq:EOM_matter_final_a} \,, \\
     \dot{H}F''(H) - \dot{\Phi}F'(H) & = -  2d \kappa^2e^{\Phi} \bar{p} \label{eq:EOM_matter_final_b} \,, \\
      2 \Ddot{\Phi} - \dot{\Phi}^2 + F(H) &=  \kappa^2 e^{\Phi} \bar{\sigma} \,,   \label{eq:EOM_matter_final_c}
    \end{align}
\end{subequations}
where a bar denotes multiplication by $\sqrt{g}=a^d$ (e.g., $\bar\rho:=\sqrt{g}\rho$), $H(t):=\dot{a}/a$ is the Hubble parameter, dots and primes denote derivatives w.r.t.~to time and $H$, respectively, and the function $F(H)$ is given by
\begin{equation}
    F(H) = 2d \sum_{k=1}^{\infty}(-\alpha')^{k-1}c_k 2^{2k} H^{2k}\,. \label{eq:F_function}
\end{equation}
These EOMs can be related via the continuity equation
\begin{equation}
    \dot{\bar\rho}+dH\bar p-\frac{1}{2}\dot\Phi\bar\sigma=0\,,
\end{equation}
which can be equivalently used to replace one of the EOMs in \eqref{eq:EOM_matter_final}.
As discussed above, these equations are invariant under the scale factor duality transformation $a\rightarrow 1/a$, since under this transformation we have
\begin{equation}\label{mattertransf}
    H\rightarrow-H\,,\quad \Phi \rightarrow \Phi\,, \quad F(H) \rightarrow F(H)\,, \quad \bar{\rho}\rightarrow \bar{\rho}\,, \quad \bar{p} \rightarrow - \bar{p}\,, \quad \bar{\sigma} \rightarrow \bar{\sigma}\,.
\end{equation}

\paragraph{A note about the symmetries}\label{sec:symmetry}
The function $F(H)$ is an even function, and it can be directly linked to the term 
\[
 \sum_{k=1}^\infty(\alpha')^{k-1}c_k\mathrm{tr}\left[(\mathcal{D}_t\mathsf{S})^{2k}\right]
\]
in the original action \eqref{eq:action_vacuum} by an overall $-1$ factor. In other words, the gravi-dilaton Lagrangian is proportional to $e^{-\Phi}[-\dot\Phi^2-F(H)]$ on FLRW when the lapse is set to unity. Thus, the symmetry $F(H)\rightarrow F(H)$ must be satisfied for the full action to be $\mathrm{O}(d,d)$ invariant.
From Eq.~\eqref{eq:F_function}, another requirement is that $F(H)$ must vanish when $H=0$, i.e., one must recover the Minkowski solution.

Notice, though, that a given solution to the EOMs might not be invariant under the scale factor duality stated above. In fact, the symmetry may well be spontaneously broken. However, the point is that the symmetry remains at the level of the full action and EOMs, so the duality always allows one to take a solution to another (different) solution \cite{Gasperini:1991qy}. Yet, since $F(H)$ has to be invariant under the duality at the level of the action, any `solution' for $F(H)$ must respect the symmetry. To be clear, string theory may provide some sets of parameters $c_k$ that determine possible expressions for $F(H)$ exactly, so $F$ is not the solution to any EOM, it is an input. As almost all the parameters $c_k$ are yet to be calculated  from string theory, the best one can do at this point is to determine whether certain solutions admit acceptable functions $F(H)$. In that sense, proposing different ans\"{a}tze for the function $F(H)$ may yield different possible background solutions, but the reverse can also be done:
given some sensible background solution, one can determine what form $F(H)$ must take and judge whether it is theoretically admissible or not.

\subsection{The different definitions of pressure}\label{sec:linktononOdd}

As it was discussed above, the EOMs are obtained after considering independent variations of the action in relation to the $\mathrm{O}(d,d)$-covariant fields $\{n,\Phi,\mathsf{S}\}$. Thus, when the action is varied in relation to $\mathsf{S}$, which encodes the spatial components of the metric, one could think that it would be equivalent to varying the action in relation to the spatial part of the metric, thus defining the pressure. However, this is generally not the case as the metric also appears implicitly in the shifted dilaton, which is kept fixed while the action is varied in relation to $\mathsf{S}$. Therefore, the pressure defined in \eqref{eq:pressure} is not the `total pressure', $p_\mathrm{t}$, which also includes the variation of the volume factor in its definition. Note that this was already partly noticed in the literature (see, e.g., \cite{Gasperini:2007ar} for more details). The difference between an $\mathrm{O}(d,d)$-covariant pressure and a diffeomorphic-covariant one, and its consequences, has also been explored in \cite{Angus:2019bqs}.

To make a connection between the different definitions of pressure, one can calculate the total pressure using the typical definition from GR or non-$\mathrm{O}(d,d)$-covariant dilaton gravity,
\begin{equation}
    p_{\mathrm{t}}=-\frac{2}{\sqrt{-G}}\frac{g^{ij}}{d}\frac{\delta S_\mathrm{m}}{\delta g^{ij}}=-\frac{2}{\sqrt{-G}}\frac{g^{ij}}{d}\left[\left(\frac{\delta S_\mathrm{m}}{\delta g^{ij}}\right)_{\Phi} + \frac{\delta S_\mathrm{m}}{\delta \Phi}\frac{\delta \Phi}{\delta g^{ij}}\right]\,.
\end{equation}
This is the definition of pressure that enters in the usual dilaton-gravity EOMs found in, e.g., \cite{Gasperini:2002bn,Gasperini:2007zz}.
Comparing it with Eqs.~\eqref{eq:dilaton_density} and \eqref{eq:pressure}, it becomes clear that
\begin{equation}
    p_{\mathrm{t}} = \frac{g^{ij}}{d} \left(T_{ij} + \frac{\sigma}{2}g_{ij}\right) = p + \frac{\sigma}{2}\,,\label{eq:ptrel}
\end{equation}
where the $T_{ij}$ in the first equality includes only the dependence on $\mathsf{g}^{-1}$ through $\mathsf{S}$, i.e., it is the spatial part of the $\mathrm{O}(d,d)$-covariant energy-momentum tensor, and so we may also call $p$ the `$\mathrm{O}(d,d)$ pressure' when there is possible confusion with the total pressure $p_\mathrm{t}$.
The importance of emphasizing the relation between $p_\mathrm{t}$ and $p$ will be seen, for instance, after considering the low-energy limit of $\alpha'$-complete cosmology in order to recover Einstein gravity.
This is discussed in the next subsection.

To gain intuition about the different definitions of pressure, let us consider the possibility of writing the matter action as
\begin{equation}
    S_\mathrm{m}[\Phi,n,\mathsf{S},\chi]=\int\mathrm{d}t\,ne^{-\Phi}\mathcal{L}_\mathrm{m}[n,\mathsf{S},\chi]\,,\label{eq:matteractiongeneral}
\end{equation}
where $\mathcal{L}_\mathrm{m}$ is the matter Langrangian density.
This is not meant to be the most general expression for the matter Lagrangian. However, there exists motivation for exploring such an expression.
By the dilaton theorem \cite{Hata:1992it,Bergman:1994qq,DiVecchia:2016szw}, one expects any explicit $\Phi$ dependence in the matter action to appear as an overall exponential factor. The matter Lagrangian could still involve couplings to derivatives of the dilaton, i.e.~dependencies of the form $\mathcal{D}_t\Phi$, but as shown in \cite{Hohm:2019jgu,Bernardo:2019bkz}, such dependencies can be removed by field redefinitions. Therefore, we argue that a matter action of the form of Eq.~\eqref{eq:matteractiongeneral} -- where we may say that matter is minimally coupled to the $\mathrm{O}(d,d)$ dilaton in the string frame -- is very generic.
Another possibility would be the case where there is strictly no $\Phi$ dependence is $S_\mathrm{m}$ (this is the case of, e.g., strings themselves \cite{Gasperini:1991ak,Gasperini:2007zz,Angus:2018mep}), in which case the dilaton charge density always vanishes.
Most of the results in string cosmology actually assume that there is no $\Phi$ dependence in the matter action,
so that matter couples only minimally to the metric in the string frame.
In such a case, this is sorting out the metric as special in relation to the two-form and dilaton fields, and the $\mathrm{O}(d,d)$ invariance is explicitly broken down to the discrete scale factor duality already at the level of the action.

An implication of \eqref{eq:matteractiongeneral} is that when the Lagrangian is independent of the spatial metric, i.e.~$\mathcal{L}_\mathrm{m}=\mathcal{L}_\mathrm{m}[n,\chi]$, it follows that $p=0$ and $p_\mathrm{t}=\sigma/2$.
An explicit example is discussed in Sec.~\ref{sec:linear_dilaton}, but let us mention another example here already: one expects a scalar field in an FLRW background that minimally couples to the metric in the string frame and that respects $\mathrm{O}(d,d)$ invariance to enter the matter action as in \eqref{eq:matteractiongeneral} with $\mathcal{L}_\mathrm{m}=\dot\chi^2/(2n^2)-V(\chi)$. It is clear that this is independent of the spatial metric, hence the $\mathrm{O}(d,d)$ pressure vanishes. The na\"ive EoS of a scalar field is only recovered through the calculation of the total pressure in such a case.
In fact, one could argue that most matter Lagrangians minimally coupled to the $\mathrm{O}(d,d)$ dilaton in the string frame, i.e.~entering the action as in \eqref{eq:matteractiongeneral}, should be independent of the spatial metric in a cosmological background. Indeed, matter fields in FLRW should just be functions of time (and thus also of the lapse) in the action; they certainly are not expected to directly depend on the scale factor. Consequently, cases where $p=0$ shall be explored with special interest.

\subsection{Low- and high-curvature limits}\label{sec:lims}

\subsubsection{Recovering Einstein gravity}\label{sec:Einstein_gravity}

Intuitively, the Einstein-gravity limit corresponds to the case where the dilaton is constant\footnote{Note that constant-dilaton solutions break the scale factor duality spontaneously.}, so that there is no distinction between the string frame and the Einstein frame, and where the $\alpha'$ corrections are negligible. This can be seen explicitly by taking the lowest-order limit in $\alpha'$ in \eqref{eq:EOM_matter_final} to get (recall $F(H)=-dH^2$ to lowest order)
\begin{subequations}\label{eq:componenteqs2}
\begin{align}
    2\dot\phi^2-2dH\dot\phi+\frac{d(d-1)}{2}H^2&=\kappa^2 e^{2\phi}\rho\,,\label{eq:NEOM0}\\
    \dot H-2H\dot\phi+dH^2&=\kappa^2 e^{2\phi}p\,,\label{eq:aEOM0}\\
    4\ddot\phi-4\dot\phi^2+4dH\dot\phi-2d\dot H-d(d+1)H^2&=\kappa^2 e^{2\phi}\sigma\,.\label{eq:phiEOM0}
\end{align}
\end{subequations}
Then, eliminating $\dot H$ and $H^2$ from Eq.~\eqref{eq:phiEOM0}, one obtains the following equation for the dilaton,
\begin{equation}
    \ddot\phi-2\dot\phi^2+dH\dot\phi=-\frac{\kappa^2e^{2\phi}}{2}\left(\rho-dp-\frac{\sigma}{2}\right)\,,\label{eq:phiEOM0e}
\end{equation}
from which it is clear that a vanishing-$\dot\phi$ solution is only possible if the right-hand side (RHS) is vanishing, i.e.~if
\begin{equation}
    \rho-dp-\frac{\sigma}{2}=0\label{eq:constantphiconstrSF0}
\end{equation}
or equivalently $\rho-dp_\mathrm{t}+(d-1)\sigma/2=0$ recalling \eqref{eq:ptrel}.
Thus, using the above,
Eqs.~\eqref{eq:componenteqs2} for a constant dilaton $\phi=\phi_0$ reduce to
\begin{equation}
    \frac{d(d-1)}{2}H^2=\tilde\kappa^2\rho\,,\qquad-(d-1)\dot H=\tilde\kappa^2(\rho+p_\mathrm{t})\,,
    \label{eq:Friedmann_eqs}
\end{equation}
where $\tilde{\kappa}^2 := \kappa^2 e^{2\phi_0}$. Providing the total pressure with an EoS parameter $w_\mathrm{t}:=p_\mathrm{t}/\rho$ and the dilatonic charge density with an EoS parameter $\lambda:=\sigma/\rho$, the constraint equation \eqref{eq:constantphiconstrSF0} can be equivalently written as
\begin{equation}
    w_\mathrm{t}=\frac{1}{d}\left(1+\frac{d-1}{2}\lambda\right)\,.\label{eq:wtlambdarel}
\end{equation}
Correspondingly, Eqs.~\eqref{eq:Friedmann_eqs} are the $(d+1)$-dimensional Friedmann equations sourced by a fluid with EoS satisfying the above, i.e.~where fixing the matter EoS amounts to fixing the dilatonic charge density or vice versa. However, we point out that one does not have the freedom to give arbitrary values to both $w_\mathrm{t}$ and $\lambda$ -- they must be related according to \eqref{eq:wtlambdarel} for a low-curvature, fixed-dilaton solution to be consistently recovered.

Therefore, for a constant dilaton and $\sigma=0$ ($\lambda=0$), but with $p\neq 0$ so that we are not in vacuum, the only consistent matter is a radiation fluid with $p/\rho=p_\mathrm{t}/\rho=1/d$ (in agreement with the known result \cite{Tseytlin:1991xk}). Thus, to have an arbitrary EoS $w_\mathrm{t}$, the dilatonic charge must be non-vanishing for a constant dilaton field when the lowest-order equations are being considered.

\paragraph*{A note on the Einstein frame}
Transforming the dilaton-gravity matter action at $0^\mathrm{th}$ order in $\alpha'$ to the Einstein frame (EF) -- cf.~Appendix \ref{app:EF} -- yields an action in the form \eqref{eq:EFaction0thorder} with the addition of a matter action $S_\mathrm{m}^{(\mathrm{EF})}$.
The resulting EOMs can be written as\footnote{We drop the subscript `E' in this paragraph; everything is meant to be in the Einstein frame. We also temporarily use units with $M_\mathrm{Pl}=1$.}
\begin{subequations}
\begin{align}
    R_{\mu\nu}-\frac{1}{2}RG_{\mu\nu}&=T_{\mu\nu}^{(\phi)}+T_{\mu\nu}^{(\mathrm{m})}\,,\\
    \Box\phi:=G^{\mu\nu}\nabla_\mu\nabla_\nu\phi&=\frac{\sigma_\phi}{4}-\frac{T^{(\mathrm{m})}}{d-1}\,,\label{eq:KGEF0}
\end{align}
\end{subequations}
where the energy-momentum tensor of the dilaton is given by
\begin{equation}
    T_{\mu\nu}^{(\phi)}=\partial_\mu\phi\partial_\nu\phi-\frac{1}{2}G_{\mu\nu}G^{\alpha\beta}\partial_\alpha\phi\partial_\beta\phi\,,
\end{equation}
and $T^{(\mathrm{m})}:=G^{\mu\nu}T_{\mu\nu}^{(\mathrm{m})}$ is the trace of the energy-momentum tensor of matter, which is defined as before, except for the Einstein-frame matter action in this case. The Einstein-frame $\phi$ charge ($\sigma_\phi$) is similarly defined as before.
From the above, it is clear that a constant-$\phi$ solution reduces to Einstein gravity. Such a solution is only obtainable, however, if the RHS of Eq.~\eqref{eq:KGEF0} vanishes, i.e.~if
\begin{equation}
    \frac{\sigma_\phi}{4}-\frac{T^{(\mathrm{m})}}{d-1}=0 \implies\frac{\sigma_\phi}{4}+\frac{\rho-dp_\mathrm{t}}{d-1}=0\,,
\end{equation}
where the second equality applies for a perfect fluid in FLRW.
This is exactly the same requirement as in the string frame [cf.~below Eq.~\eqref{eq:constantphiconstrSF0}], recalling that equations of state should remain frame invariant and that $\sigma_\phi=2\sigma$ (which simply follows from the fact that $\sigma$ is the variation of the matter action w.r.t.~$\Phi$, while $\sigma_\phi$ is the variation w.r.t.~$\phi$, and one has $\delta\Phi=2\delta\phi$).
In particular, when $\sigma_\phi=0$, the constraint is $\rho-dp_\mathrm{t}=0$, which is the EoS of radiation.
When this requirement is met in the matter sector, the dilaton EOM simply becomes $\Box\phi=0$, or $\ddot\phi+dH\dot\phi=0$ in FLRW, i.e., it is the Klein-Gordon equation of a massless scalar field. It is clear that there are two solutions: $\dot\phi=0$, from which Einstein gravity immediately emerges, and $\dot\phi\propto a^{-d}$. For the latter, note that $\dot\phi\propto a^{-d}\rightarrow 0$ as $a\rightarrow\infty$, so Einstein gravity is recovered asymptotically in the small-curvature limit, as expected. Note that something similar happens when looking at the Einstein-gravity limit of other modifications to GR that involve a scalar field non-minimally coupled to gravity. For instance, in Brans-Dicke gravity, Einstein gravity with the addition of a massless scalar field is recovered when the trace of the matter energy-momentum tensor vanishes (e.g., as is the case for radiation); see, e.g., \cite{Banerjee:1996iy,Faraoni:2018nql,Faraoni:2019sxw}.

\paragraph{To all orders in $\alpha'$}\label{par:GRalphaprime}
Back to the EOMs \eqref{eq:EOM_matter_final}, one can define $A(H):=F(H)+dH^2$, so that using Eq.~\eqref{eq:F_function},
\begin{equation}
    A(H)=2dH^2\sum_{k=1}^\infty(-1)^k2^{2(k+1)}c_{k+1}(\alpha'H^2)^k\,.\label{eq:defA}
\end{equation}
Thus, $A(H)$ represents all the $\alpha'$ corrections, starting at order $(\alpha')^1$, i.e.~pulling out the expected lowest-order contribution from $F(H)$.
Accordingly, $F'=-2dH+A'$ and $F''=-2d+A''$, so the EOMs \eqref{eq:EOM_matter_final} read
\begin{subequations}\label{eq:OddwA}
\begin{align}
    \dot\Phi^2-dH^2+HA'-A&=2\kappa^2e^\Phi\bar\rho\,,\label{eq:OddFriedwA}\\
    -2d\dot H+\dot HA''+2dH\dot\Phi-\dot\Phi A'&=-2d\kappa^2e^\Phi\bar p\,,\label{eq:OdddotHwA}\\
    2\ddot\Phi-\dot\Phi^2-dH^2+A&=\kappa^2e^\Phi\bar\sigma\,.\label{eq:OddPhiddwA}
\end{align}
\end{subequations}
To explore the Einstein-gravity limit, consider the ansatz\footnote{Note that it is not possible here to isolate $H^2$ from Eq.~\eqref{eq:OddFriedwA} and correspondingly obtain an equation for $\phi$ in the form of Eq.~\eqref{eq:phiEOM0e} as it was possible for the $0^\mathrm{th}$ order in $\alpha'$. Thus, it is not possible to write down a simple constraint equation on the matter sector for the solution $\dot\phi=0$ to be obtainable. Nevertheless, it is assumed that the solutions can be obtained and the corresponding consequences are explored.} $\dot\phi=0$ (say $\phi=\phi_0$), which implies $\dot\Phi=-dH$, hence the EOMs become
\begin{subequations}
\begin{align}
    \frac{d(d-1)}{2}H^2+\frac{1}{2}(HA'-A)&=\tilde\kappa^2\rho\,,\label{eq:OddphiconstFried}\\
    \dot H+dH^2-\frac{1}{2d}\left(\dot HA''+dHA'\right)&=\tilde\kappa^2p\,,\\
    \dot H+\frac{d+1}{2}H^2-\frac{1}{2d}A&=-\frac{1}{2d}\tilde\kappa^2\sigma\,,\label{eq:OddphiconstPhiEOM}
\end{align}
\end{subequations}
where all the terms involving $A(H)$ represent the $\alpha'$ corrections to the Friedmann equations starting at $\mathcal{O}(\alpha'H^4)$.
Assume now there is a solution $H(t)$, which at a given moment in time satisfies
\begin{equation}
    H = b M_\mathrm{s}\label{eq:Hubble_b_parameter}\,,
\end{equation}
where $M_\mathrm{s} := 1/\ell_\mathrm{s} \sim 1/\sqrt{\alpha'}$ is the string mass and $b$ is some positive constant.
If $b \ll 1$, it corresponds to a low-curvature regime.
Then, focusing on the modified Friedmann constraint equation \eqref{eq:OddphiconstFried}, the $H^2$ term in the LHS scales as $b^2 M_\mathrm{s}^2$, while the `new' contribution scales as $HA'-A\sim A\sim\alpha'H^4+\mathcal{O}(\alpha'^2H^6)\sim b^4 M_\mathrm{s}^2+\mathcal{O}(b^6M_\mathrm{s}^2)$.
Clearly, this is much smaller than the $H^2$ contribution by a factor of $b^2 \ll 1$, therefore it cannot contribute much.
Rather, it must be the RHS, $\tilde\kappa^2\rho$, that scales as $H^2 \sim b^2 M_\mathrm{s}^2$.
Here, $\tilde\kappa^2\sim M_\mathrm{Pl}^{1-d}$ for a fixed dilaton at weak coupling, so $\rho$ must scale as
$b^2 M_\mathrm{s}^2 M_\mathrm{Pl}^{d-1}$.
This can also be written as $\rho\sim b^2 (M_\mathrm{s}/M_\mathrm{Pl})^2 M_\mathrm{Pl}^{d+1}$, where in the weak-coupling regime, one has $M_\mathrm{s}/M_\mathrm{Pl}=e^{2\phi_0}\ll 1$, so $\rho$ is much smaller than the Planck density $M_\mathrm{Pl}^{d+1}$.
Still, the matter source
dominates over the higher $\alpha'$ corrections in the weak-curvature regime.
This makes sense: when $H$ is small, the $\alpha'$ corrections are very small, and $H^2\sim\tilde\kappa^2\rho$, so most matter contents actually dominate over curvature corrections. In that sense, one can say that `Einstein gravity is recovered.'

\subsubsection{The high-energy limit}

Let us consider the high-energy limit and the role that matter plays at such scales. In a high-curvature regime where $H\sim M_\mathrm{s}$ [so $b\sim\mathcal{O}(1)$ in Eq.~\eqref{eq:Hubble_b_parameter}], one has $A(H)\sim M_\mathrm{s}^2$ (to all orders in $\alpha'$), and it looks like it is rather the $\alpha'$ corrections that dominate over matter, unless $\tilde\kappa^2\rho$ is also of order $M_\mathrm{s}^2$.
However, this can be the case if $\rho\sim M_\mathrm{s}^2 M_\mathrm{Pl}^{d-1} = (M_\mathrm{s}/M_\mathrm{Pl})^2 M_\mathrm{Pl}^{d+1}\ll M_\mathrm{Pl}^{d+1}$ in the weak-coupling regime.
Thus, even in the sub-Planckian density regime, matter can be as important as the non-perturbative $\alpha'$ corrections when the curvature scale is the string scale.
Correspondingly, exploring the role of matter in high-curvature regimes such as in the very early universe is well justified.
This is further developed in the second part of the paper.

In a high-scale regime and for duality preserving solutions (as motivated by examples of stringy early universe cosmologies such as string gas cosmology and pre-big bang cosmology), constant-dilaton solutions are not expected since their $\mathrm{O}(d,d)$ transformation has a time-dependent dilaton for $H(t) \neq \mathrm{const.}$, spontaneously breaking the $\mathrm{O}(d,d)$ invariance. In the context of very early universe cosmology, one should check case by case whether the solutions have some `graceful exit mechanism', i.e.~moving away from the high-curvature regime into a small-curvature regime corresponding to, e.g., the onset of radiation domination and onward. Moreover, the dilaton should\footnote{Actually, it may be that the dilaton does not have to freeze completely. A time-dependent dilaton in late-time cosmology could still be consistent with observations and various constraints, provided how much the dilaton couples to everything else satisfies certain bounds (see, e.g., \cite{Brax:2019rwf}).} freeze (i.e., reach the `Einstein-gravity limit' with $\dot\phi=0$), wishfully dynamically, although there could exist other mechanisms enabling the dilaton to freeze.\footnote{For instance, there could be some condensation mechanism or some symmetry breaking coming from some dilaton potential. Some other form of dilaton potential could also presumably allow for a dynamical freeze out of the dilaton, see \cite{Damour:1994ya,Damour:1994zq}.}
The present paper shall not be concerned though with the construction of full cosmological scenarios. This has been first explored in, e.g., \cite{Bernardo:2020nol,Bernardo:2020bpa} and deserves to be extended in future work.

\subsection{Recovering the linear dilaton CFT}\label{sec:linear_dilaton}

In this section, a well-known result related with the cosmological constant case is recovered: the linear dilaton CFT, with $\{G_{\mu\nu}, B_{\mu\nu}, \phi\} = \{\eta_{\mu\nu}, 0, V_{\mu}x^\mu\}$, where $V^{\mu}$ is a constant vector. This solution should hold to all orders in $\alpha'$ because the dilaton is a linear function of the worldsheet embedding coordinates, and so its coupling in the Polyakov action is such that the path integral is Gaussian after a shift in variables.\footnote{This had already been noticed in \cite{Gasperini:1996fu}. They arrived at the same conclusion by analyzing what would be the corresponding shape of the EOMs in the presence of all $\alpha'$ corrections in a background with constant curvature and a linear dilaton, and then they inferred such a background would be a solution.}
If one works in arbitrary $D$ dimensions, then the spacetime action for the string background fields contains a term \cite{Polyakov:1981rd, Callan:1985ia}
\begin{equation}
    S_\mathrm{m} = \frac{1}{2\kappa^2}\int\mathrm{d}^Dx\,\sqrt{-G} e^{-2\phi}\left(-\frac{2}{3\alpha'}(D-D_\mathrm{c})\right)=-\frac{1}{2\kappa^2}\int\mathrm{d}^Dx\,n e^{-\Phi} \Lambda\,,
\end{equation}
with $\Lambda := 2(D-D_\mathrm{c})/(3\alpha')$, where $D_\mathrm{c}$ is the critical dimension of the string theory considered. If $D \neq D_\mathrm{c}$, such a term in the action is mandatory for consistency with the vanishing of the $\beta_{\phi}$ functional. This term enters as a `cosmological constant' in the string-frame action as it has the same dilaton-dependent overall factor as the Ricci scalar term associated to the string metric. As such, it is a genuine cosmological-constant term when the dilaton is constant.

Upon assuming the full metric to be flat (so in particular $H\equiv 0$), the EOMs \eqref{eq:EOM_matter_final} reduce to
\begin{subequations}\label{eq:EOMcc}
\begin{align}
    \dot{\Phi}^2 &= 2\kappa^2 e^{\Phi} \rho = \Lambda\,,\\
        2\ddot{\Phi} - \dot{\Phi}^2 &= \kappa^2 e^{\Phi}\sigma =-\Lambda\,,
\end{align}
\end{subequations}
where the energy density, the dilatonic charge and the $\mathrm{O}(d,d)$ pressure were evaluated for the above matter action. Note that the $\mathrm{O}(d,d)$ pressure vanishes here, and the pressure equation \eqref{eq:EOM_matter_final_b} is trivially satisfied in this case. This is an interesting example where the matter action is of the form of \eqref{eq:matteractiongeneral} and where the presence of dilatonic charge implies two different definitions of pressure. The $\mathrm{O}(d,d)$ pressure that naturally enters the EOMs is not the expected one with EoS $p=-\rho$ for a positive cosmological constant. Rather, one has $p=0$, and it is the total pressure which satisfies $p_\mathrm{t}=p+\sigma/2=-e^{-\Phi}\Lambda/(2\kappa^2)=-\rho$. The solution to \eqref{eq:EOMcc} for $\Phi(t)$ is
\begin{equation}
    \Phi(t) = 2\phi(t) = \sqrt{\Lambda}\,t + \mathrm{constant}\,,
\end{equation}
which is exactly of the form $\phi(t) = V_\mu x^\mu$ for $V^{\mu} =- \sqrt{\Lambda/4}\,\delta^{\mu}_{\;0}$ (the constant can be set to zero by changing the origin of time). For consistency, $\Lambda>0$ and so $D>D_\mathrm{c}$. This is compatible with the fact that a purely time-dependent linear dilaton of the form $\phi(t) = V_\mu x^\mu$ should have a time-like derivative vector $\nabla_\mu\phi=V_\mu$. Thus, the linear dilaton CFT solution is recovered for the fully $\alpha'$-corrected equations, a successful test of the framework, which had not been previously appreciated. 

From Eqs.~\eqref{eq:tEFtransf} and \eqref{eq:HEFtransf} in Appendix \ref{app:EF}, the Einstein-frame evolution corresponding to the above solution is
\begin{equation}
    H_\mathrm{E}=\frac{1}{t_\mathrm{E}}\,,
\end{equation}
which corresponds to a solution of the $(d+1)$-dimensional Friedmann equations with effective EoS $w_\mathrm{E}=(2-d)/d$, thus saturating the SEC (we expand on this topic in Sec.~\ref{sec:SEC}). Indeed, the above is equivalent to $a_\mathrm{E}\propto |t_\mathrm{E}|$.
Therefore, the non-critical static background in the string frame with running dilaton corresponds to non-accelerating expansion in the Einstein frame (though non-decelerating either, i.e., it is precisely on the margin).

\subsection{A note about constant-dilaton solutions}\label{sec:constantdilaton}

The linear dilaton CFT background in the last subsection is an example of a solution with a trivial string-frame metric (Minkowski) that translates to a non-trivial Einstein-frame cosmology due to the running of the dilaton. One immediate question is whether there are solutions with constant dilaton such that there is no difference between the frames.
In the following, it is shown that consistent solutions with constant dilaton and vanishing $\mathrm{O}(d,d)$ pressure\footnote{We recall that $p=0$ is well motivated for most matter Lagrangian densities that minimally couple to the string-frame metric (cf.~Sec.~\ref{sec:linktononOdd}) and that this is in most cases not equivalent to having pressureless matter with $p_\mathrm{t}=0$.} have $H(t)\neq \mathrm{constant}$ if and only if $\rho \neq -\sigma/2$, and the only solutions in vacuum with constant dilaton are Minkowski or dS ones. To show this, let us first set the dilaton to be constant in the EOMs \eqref{eq:EOM_matter_final}, yielding
\begin{subequations}\label{constantdilaton}
\begin{align}\label{constantdilatonrho1}
    d^2H^2 + H F'(H) - F(H) &= 2\tilde{\kappa}^2 \rho\,,\\
    \dot{H}F''(H) + d HF'(H) &= -2d\tilde{\kappa}^2 p \,,\label{constantdilatonp1}\\
    -2d\dot{H} - d^2H^2 + F(H) &= \tilde{\kappa}^2 \sigma\,,\label{constantdilatonsigma1}
\end{align}
\end{subequations}
where as denoted before $\tilde{\kappa}^2 = \kappa^2 e^{2\phi_0}$.
Adding the first and third equations and eliminating $\dot H$ using the second equation, we arrive at
\begin{equation}
    (F''(H)+2d^2)HF'(H)=2\tilde\kappa^2\left[\left(\rho+\frac{\sigma}{2}\right)F''(H)-2d^2p\right]\,.
\end{equation}
If we assume $\rho+\sigma/2=0$ and $p=0$, then possible solutions are found only if $F''(H)=-2d^2$, $H=0$, or $F'(H)=0$. In every case, this is only possible if the Hubble parameter is a constant, $H=H_0$.
Upon reinserting this in \eqref{constantdilaton}, we arrive at the following consistency relations:
\begin{subequations}
\begin{align}
    d^2H_0^2-F(H_0)&=2\tilde\kappa^2(\rho+p)\,,\\
    H_0F'(H_0)&=2\tilde\kappa^2\left(\rho+\frac{\sigma}{2}\right)=-2\tilde\kappa^2 p\,.
\end{align}
\end{subequations}
If we assume $p=0$ again, it implies that $\rho+\sigma/2=0$ must follow by consistency, and furthermore, one recovers $H_0F'(H_0)=0$, whose solutions are $H_0=0$ or $F'(H_0)=0$, together with the additional constraint that $F(H_0)=d^2H_0^2-2\tilde\kappa^2\rho$.

So to summarize: if we assume a constant dilaton ($\dot\phi=0$) and vanishing $\mathrm{O}(d,d)$ pressure ($p=0$), then the equations are only consistent if the matter sector satisfies $\rho=-\sigma/2$, which is equivalent to a total pressure $p_{\mathrm{t}}= p+\sigma/2=-\rho$. This is possible with a cosmological constant EoS (this solution was first found in \cite{Bernardo:2019bkz}) or in vacuum ($p=\rho=\sigma=0$). The resulting allowed solutions are either Minkowski with $H=0$ (in vacuum) or dS with $H=H_0\neq 0$ (in vacuum or with matter satisfying the aforementioned EoS) as long as $F'(H_0)=0$ and $F(H_0)=d^2H_0^2-2\tilde\kappa^2\rho$. Put differently, if we assume a vacuum, then the only allowed solutions are Minkowski space or dS space. Note that, in the latter case, one really means exact dS spacetime since the dilaton is constant, and hence, there is no distinction between the string frame and the Einstein frame. While the fact that adding matter with EoS $p_\mathrm{t}=-\rho$ to the system yields a dS solution is not a surprise\footnote{The discussion above applies directly to the cosmological constant case considered in the last subsection, since $w=0$ and $\lambda = -2$ for that case, i.e., there is also a solution with constant dilaton and constant Hubble parameter that is a dS solution sourced by a cosmological constant. However, as discussed in \cite{Bernardo:2020zlc}, it is worth mentioning that there can be non-trivial dS solutions with $w=0$ and $\lambda=-2$, although with running dilaton.}, finding a dS solution in vacuum and in the Einstein frame\footnote{In \cite{Hohm:2019jgu}, an argument against vacuum dS solutions with constant dilaton was developed. However, it was assumed that $F'(H_0) \neq 0$, which is precisely the condition for having the dS solution above. There is \textit{a priori} no reason not to allow $F'(H_0)=0$ as a possibility.} is quite unusual, though this was first discovered in \cite{Nunez:2020hxx}.

Generically, higher-derivative corrections in an action may bring extra degrees of freedom into the dynamics since extra initial conditions would be required to solve the higher-order EOMs. However, in the $\alpha'$-complete cosmology framework, the EOMs are still second order, and there should not be extra propagating degrees of freedom compared with the lowest-order theory.\footnote{In the context of effective field theory approaches to modifying GR, higher-curvature corrections that yield at most second-order EOMs on an FLRW background have been explored in \cite{Cano:2020oaa}. An interesting observation is made: the resulting theory is modified in a very similar fashion to what is found in the string theory context of this paper with an infinite tower of $\alpha'$ corrections. It is also noticed that new dS solutions are allowed (in particular in vacuum) when the Hamiltonian constraint (which is an infinite series in $H$) has non-trivial roots. As is pointed out though, such a solution remains at the level of a toy model (it can only be exact dS); in particular, it cannot continuously evolve toward a low-energy GR-like regime.} This fact also matches what is expected from string theory, as the $\alpha'$-corrected spacetime theory should have only the graviton, dilaton, and Kalb-Ramond fields in the universal massless spectrum. After turning off the matter sector and imposing a constant dilaton, the dynamics for a FLRW background is determined by an equation for $a(t)$, with solution space fully constrained by the Hamiltonian constraint (the EOM for the lapse function).
The equation for $a(t)$ \eqref{constantdilatonp1} fixes $H$ to be a constant, and in this case the tower of $\alpha'$ corrections in the action may act as an effective cosmological constant. At lowest order, $F(H) = -dH^2<0$, and the only solution to the constraint \eqref{constantdilatonrho1} is a Minkowski solution, but if the function $F'(H)$ has non-trivial roots $H_0^{(i)}$, such that $F(H_0^{(i)})=d^2(H_0^{(i)})^2>0$, there are non-perturbative dS solutions for the constraint. Since all $\alpha'$ corrections contribute to the effective cosmological constant, $H_0^{(i)}$ is expected to be at the string scale, with $\sqrt{\alpha'} H_0^{(i)} \sim 1$, thus avoiding the no-go theorem in \cite{Kutasov:2015eba}. More comments on this shall be given in the discussion section.

\section{Effective energy conditions and their violation}\label{sec:SEC_disc}

The singularity theorems of GR \cite{ Penrose:1964wq,Hawking:1967ju,Hawking:1969sw} ensure the incompleteness of geodesics in spacetimes that satisfy reasonable assumptions.
In particular, the strong energy condition (SEC) is assumed for proving timelike incompleteness in cosmological contexts, while the null energy condition (NEC) is required for the proof of null incompleteness in gravitational collapsing spacetimes. In the proof of such theorems, the Einstein field equations are used to connect purely geometric quantities (expansion of null or timelike geodesic congruences) with properties of the matter energy-momentum tensor. Thus, there is no difference in asserting that the energy-momentum tensor $T_{\mu\nu}$ or the Ricci tensor $R_{\mu\nu}$ satisfies a given energy condition, insofar as the Einstein equations are considered.

However, if the gravity action contains higher-order terms, the Einstein equations are modified and then one needs to be careful with what a given energy condition implies. It is still possible to use the singularity theorems provided that the energy conditions are seen as a geometrical condition to be applied on the Ricci tensor. Operationally, all modifications to the Einstein equations are added to the matter side of the equations such that the system is described by GR sourced by an effective energy-momentum tensor $T_{\mu\nu}^{\mathrm{eff}}$. Thus, geometric conditions on the Ricci tensor are equivalent to energy conditions on $T_{\mu\nu}^{\mathrm{eff}}$. It is in this sense that the energy conditions are discussed in this section. Moreover, since parallel transport depends on frame, geodesic incompleteness and the singularity theorems are not invariant under metric field redefinitions, such as the one used to switch from the string frame to the Einstein frame (and vice versa). In this section, conditions are applied to the Ricci curvature of the Einstein-frame metric, $R_{\mu\nu}^\mathrm{E}:=R_{\mu\nu}[\bm{G}_\mathrm{E}]$, such that the GR theorems' results can be promptly applied.

The discussion above is relevant for constructing potentially non-singular solutions from the $\alpha'$-corrected equations considered in the previous sections. For cosmological spacetimes, violation of the SEC is required for accelerating solutions, while non-singular bouncing cosmologies should violate the NEC during the bouncing phase. Hence, investigating violations of these energy conditions in $\alpha'$-complete cosmology can help establish accelerating and/or bouncing non-perturbative solutions. From the string theory perspective, these solutions would avoid supergravity no-go theorems \cite{Maldacena:2000mw,Gibbons:2003gb} by means of the tower of duality-invariant $\alpha'$ corrections.

\subsection{SEC violation and accelerating solutions}\label{sec:SEC}

The SEC states that $R_{\mu\nu}^\mathrm{E}u^\mu u^\nu \geq 0$ at every spacetime point for any timelike vector $u^\mu$. Physically, this ensures that any observer would locally measure gravity to be `attractive', in the sense that timelike geodesics will be focused due to the spacetime curvature. For a perfect fluid satisfying the SEC, comoving observers in a cosmological background would measure $(d-2)\rho_\mathrm{total} + dp_{\mathrm{total}} \geq 0$ and $\rho_{\mathrm{total}} + p_{\mathrm{total}} \geq 0$.
In particular, an EoS $p_\mathrm{total}<(2-d)\rho_\mathrm{total}/d$ violates the SEC, and one notices that this is precisely the condition for a perfect fluid to yield an accelerating background solution in GR. As will become clear below, SEC violation may thus be used as a criterion for finding accelerating solutions in the Einstein frame.

For a flat $(d+1)$-dimensional FLRW background in the Einstein frame and for comoving observers with $u^\mu = (1,\vec{0})$, the SEC reduces to
\begin{equation}\label{strongenergyconditionatansatz}
    R_{00}^\mathrm{E} = - d \left(\frac{\dd H_{\mathrm{E}}}{\dd t_{\mathrm{E}}} + H_{\mathrm{E}}^2\right) \geq 0\,,
\end{equation}
and using Eqs.~\eqref{eq:HEFtransf} and \eqref{eq:HEprimegentransf} of Appendix \ref{app:EF}, this further implies
\begin{equation}\label{R00beforeeq}
    R_{00}^\mathrm{E} = \frac{d e^{\frac{4\phi}{d-1}}}{d-1}\left(\ddot{\Phi} + \dot{H} + H(\dot{\Phi} + H)\right)\, \geq 0
\end{equation}
in terms of string-frame quantities.
Thus, in order to violate the SEC, we need string-frame solutions satisfying
\begin{equation}\label{secviolation}
    \ddot{\Phi} + \dot{H} + H(\dot{\Phi} + H) < 0
\end{equation}
at some time $t$. Note that, according to \eqref{eq:appEFtranf}, the above is completely equivalent to the condition $\dd^2a_\mathrm{E}/\dd t_\mathrm{E}^2>0$, which makes the connection between SEC violation and accelerated expansion in the Einstein frame explicit. Using the EOMs \eqref{eq:OddwA} to write
\begin{subequations}\label{eq:dynamical_system}
\begin{align}
    \ddot{\Phi} &= \frac{1}{2}\left(\kappa^2 e^{\Phi}\bar{\sigma}+\dot{\Phi}^2+dH^2-A(H)\right)\,,\\
    \dot{H} &= \left(1-\frac{A''(H)}{2d}\right)^{-1}\left[\kappa^2 e^{\Phi} \bar{p} + \dot{\Phi}\left(H - \frac{A'(H)}{2d}\right)\right]\,,\label{eq:dotHEOM}
\end{align}
\end{subequations}
which holds as long as $A''(H)\neq 2d$ [i.e., $F''(H) \neq 0$]\footnote{As a series expansion, one expects from \eqref{eq:defA} $A''(H)\propto \alpha' H^2+\ldots$, with a proportionality constant of at most $\mathcal{O}(1)$, and $\alpha' H^2$ is also expected to be at most $\mathcal{O}(1)$ if $H$ ends up being bounded.
Thus, it appears safe to assume $1-A''/(2d)>0$. Of course, the assumption should always be checked for a given model.},
SEC violation can be equivalently written as
\begin{equation}\label{secviolationaftereq}
    (\dot{\Phi} + H)^2 +\kappa^2e^{\Phi}\left(\bar{p} + \frac{\bar{\sigma}}{2}- \bar{\rho}\right) - A(H)+ \frac{H}{2}A'(H)+ \frac{e^{\Phi}}{2d}\frac{\mathrm{d}}{\mathrm{d}t}\left(e^{-\Phi}A'(H)\right)\,< 0\,,
\end{equation}
which is a condition on the matter sector (encoded in $\sigma$, $\rho$ and $p$) and on the $\alpha'$ corrections [encoded in $A(H)$].

A couple of checks for Eqs.~\eqref{secviolation} and \eqref{secviolationaftereq} can be done. First, for a constant dilaton $\phi=\phi_0$, one has $\dot{\Phi} = - dH$, and Eq.~\eqref{R00beforeeq} becomes
\begin{equation}
    R_{00}^\mathrm{E} = -d e^{\frac{4\phi_0}{d-1}}\left(\dot{H} + H^2\right)\,,
\end{equation}
which has the same structure as Eq.~\eqref{strongenergyconditionatansatz}. This should be so, since for a constant dilaton, the Einstein- and string-frame variables are proportionally related (in fact, the exponential factor in the equation above can be set to one by a redefinition of $t$). Secondly, considering the vacuum case and turning off $\alpha'$ corrections, the SEC cannot be violated. This is immediately seen from Eq.~\eqref{secviolationaftereq} upon setting $A(H)\equiv 0$ and $\rho=p=\sigma=0$, which yields the always-positive term $(\dot\Phi+H)^2$ on the LHS.
This should be so as, \emph{in this specific case}, the system consists of a metric and a massless scalar field in the Einstein frame.

To understand how different matter contents can allow for SEC violation and thus accelerated expansion in the Einstein frame, let us consider the set of solutions found in \cite{Bernardo:2020zlc}
of the form 
\begin{equation}
    H(t) = H_0\,, \qquad \dot{\Phi} = \frac{dw}{1+\lambda/2} H_0\,,\label{eq:H0varPhisol}
\end{equation} 
with a positive, non-trivial, constant Hubble parameter $H_0$, and where the equations of state in the matter sector are specified as $w:=p/\rho$ and $\lambda:=\sigma/\rho\neq -2$. For this class of solutions, violation of the SEC, Eq.~\eqref{secviolation}, is equivalent to\footnote{Note that the case where $w=-(1+\lambda/2)/d$ corresponds to saturating the SEC, and from \eqref{eq:HEFtransf}, it corresponds to Minkowski space in the Einstein frame. This is the basis of the loitering solution studied in \cite{Bernardo:2020nol,Bernardo:2020bpa,Jonas:2021xkx}.}
\begin{equation}
    w<-\frac{1}{d}\left(1+\frac{\lambda}{2}\right)\,,
\end{equation}
assuming $\lambda>-2$.
Note that for simple Lagrangians as discussed below \eqref{eq:matteractiongeneral} for which $p=0$, the above can never\footnote{The only exception is when $\lambda=-2$, in which case it can be shown \cite{Bernardo:2020zlc} that the solution $H=H_0$ and $\dot\Phi=-\beta H_0$ is stable for any constant $\beta>0$. Then, violating the SEC is possible whenever $\beta>1$. An example of matter with $w=0$ and $\lambda=-2$ is the cosmological constant coupling to the dilaton as discussed in Sec.~\ref{sec:linear_dilaton}.} be met, i.e., the SEC is always satisfied, and thus, accelerated expansion in the Einstein frame can never be achieved. This is an intriguing result already; for example, a simple, general scalar field that minimally couples to the dilaton in the string frame admits constant-$H$ solutions in the string frame with running dilaton according to \eqref{eq:H0varPhisol}, but it does not admit any accelerating solution in the Einstein frame.
More generally, recalling that we can write $w_\mathrm{t}=w+\lambda/2$ from $p_\mathrm{t}=p+\sigma/2$ and $w_\mathrm{t}:=p_\mathrm{t}/\rho$, the above is also equivalent to
\begin{equation}
    w_\mathrm{t}<-\frac{1}{d}+\frac{(d-1)}{2d}\lambda\,.
\end{equation}
From this, one recovers the result that the SEC is violated provided $w_\mathrm{t} < (2-d)/d$ when $\lambda = -2(d-3)/(d-1)$, which vanishes in $d=3$ spatial dimensions.
Furthermore, whenever the matter \emph{does} couple to the dilaton in the string frame (i.e.~$\lambda\neq 0$), the condition can be either relaxed or strengthened. As an example, if, say, $\lambda=2$, then SEC violation demands $w<-2/d$ or equivalently $w_\mathrm{t}<1-2/d$. In such a case, the condition on the total pressure is quite weak; in particular, it can be positive, which is usually much more easily achieved for standard matter. However, we do not know of a specific simple example which would yield equations of state satisfying the above conditions.

\subsection{NEC violation and non-singular solutions}\label{sec:NEC}

The NEC states that $R^\mathrm{E}_{\mu\nu}k^\mu k^\nu\geq 0$ at every spacetime point for any null vector $k^\mu$. For a perfect fluid in the Einstein frame and considering the same ansatz we considered for the SEC, this means that the whole matter sector, which also includes the dilaton field in this frame, has to satisfy $\rho_\mathrm{total}+p_\mathrm{total}\geq 0$. Looking at \eqref{eq:Friedmann_eqs}, one can see that this implies\footnote{Despite \eqref{eq:Friedmann_eqs} being derived for a constant dilaton, it still holds for a dynamical dilaton as long as its contribution to the pressure and energy density are also taken into account, thus defining an effective perfect fluid.}
\begin{equation}
    \frac{\mathrm{d}H_\mathrm{E}}{\mathrm{d}t_\mathrm{E}}\leq 0\,.
\end{equation}
Thus, NEC violation happens whenever the Einstein frame's Hubble parameter grows in time. Note that NEC violation implies SEC violation [cf.~Eq.~\eqref{strongenergyconditionatansatz}]. 

Using Eq.~\eqref{eq:HEFtransf}, the condition for violating the NEC can be written in terms of the string-frame quantities as
\begin{equation}
 \ddot\Phi+\dot H+(\dot\Phi+H)\left(H+\frac{\dot\Phi+H}{d-1}\right)<0\,,\label{eq:NECvcond}
\end{equation}
and it is worth noting that saturating the NEC leads to the condition for a dS solution in the Einstein frame ($H_\mathrm{E}=\mathrm{constant}$). One can see that violating the NEC is more stringent than violating the SEC since, in comparison to Eq.~\eqref{secviolation}, the above condition can be written as
\begin{equation}
    \ddot\Phi+\dot H+H(\dot\Phi+H)<-\frac{(\dot\Phi+H)^2}{d-1}<0\,.
\end{equation}
Let us consider the condition for violating the NEC \eqref{eq:NECvcond} written in terms of the $\alpha'$ corrections, the matter variables and the standard dilaton. By using Eqs.~\eqref{eq:OddFriedwA}--\eqref{eq:OddPhiddwA} to eliminate $H$, $\dot H$, and $\ddot\Phi$ in \eqref{eq:NECvcond}, and Eq.~\eqref{eq:shifteddilatondef} to rewrite everything in terms of $\phi$, the condition for violating the NEC is given by 
\begin{align}\label{eq:NECviolating_matter&corrections}
    \frac{1}{1-\frac{A''}{2d}}&\bigg((d-1)\kappa^2e^{2\phi}\left(\rho+p+\frac{1}{2}\sigma\right)+4\dot\phi^2-\frac{A''}{8d^2}\left(4A+\frac{d+1}{d-1}A'^2\right)\nonumber\\
    &\ -\frac{1}{4d}\kappa^2e^{2\phi}A''\left(2(d+1)\rho+(d-1)\sigma\right)\pm\frac{d+1}{8d^2(d-1)}\sqrt{\Delta}(A'-4\dot\phi)A''\nonumber\\
    &\ +\frac{\dot\phi}{2d^2}\left[A'\left(\frac{(d+1)^2}{d-1}A''-2d(d-1)\right)-4d\frac{d+1}{d-1}\dot\phi A''\right]\bigg)<0\,,
\end{align}
where the $\pm$ sign depends on the branch solution taken for $H$, and
\begin{equation}
    \Delta := 4d\left(4\dot\phi^2+2(d-1)\kappa^2e^{2\phi}\rho+(d-1)A\right)+A'\left(A'-8d\dot\phi\right)\,.
\end{equation}
To connect the above condition with the Einstein-gravity limit (cf.~Sec.~\ref{sec:Einstein_gravity}), one can first ignore all the $\alpha'$ corrections by imposing $A(H)\equiv 0$ in Eq.~\eqref{eq:NECviolating_matter&corrections}, resulting in
\begin{equation}
    \rho+p_\mathrm{t}<-\frac{4}{d-1}\kappa^{-2}e^{-2\phi}\dot\phi^2<0\,.\label{eq:NECvnoalphap}
\end{equation}
From this, we see that an evolving dilaton makes the condition stronger, i.e., it is harder to violate the NEC the larger the rate at which the dilaton is running. Conversely, if the dilaton is constant, the condition reduces to $\rho+p_\mathrm{t}<0$, as in standard GR (as expected).

From the above, a straightforward way to avoid the singularity theorems and potentially to have a non-singular solution is to have solutions satisfying \eqref{eq:NECvcond} or equivalently \eqref{eq:NECviolating_matter&corrections} at least for some time. Note that `non-singular' refers to the behaviour in the Einstein frame, so the solution can still have a singular behaviour in the string frame -- the quest for non-singular solutions is only well defined once the frame in which this property holds is properly specified.\footnote{Some works in the literature (e.g., \cite{Domenech:2019syf,Wetterich:2020oyy,Casadio:2020zmn}) suggest that spacetime singularities in GR may be removable thanks to field redefinitions (including frame transformations). While this appears formally true, we believe it remains an issue of determining in which frame matter as we observe it today couples minimally to the metric, or put differently, what are the actual fields at play (i.e., which redefinition) that correspond to the observed physical processes in the universe.}
We will now turn to the question of trying to find actual solutions that could effectively violate the NEC and be non-singular.

\section{Bouncing cosmology}\label{sec:bouncing}

\subsection{General requirements}\label{sec:bouncegen}

In this section, we specialize to explore the possibility of finding non-singular bouncing solutions. These solutions are particularly relevant for the early universe as they could potentially explain how the spectrum of cosmological fluctuations observed today were produced (e.g., \cite{Wands:1998yp,Finelli:2001sr,Khoury:2001wf,Lehners:2007ac,Buchbinder:2007ad,Lehners:2008vx,Ijjas:2014fja,Brandenberger:1988aj,Nayeri:2005ck,Brandenberger:2006vv,Brandenberger:2008nx}), thus representing alternatives to inflation. It is known that bouncing solutions in GR are generically singular in a flat FLRW background, so the existence of new physics is typically assumed (e.g., \cite{Cai:2007qw,Cai:2008qw,Lin:2010pf,Cai:2012va,Ijjas:2016tpn,Cai:2017dyi,Boruah:2018pvq,Mironov:2019mye,Sakakihara:2020rdy,Ilyas:2020qja} and more references therein) to account for the non-singular behavior of the universe. In our context, the hope is that this new physics is given by the infinite tower of $\alpha'$ corrections as motivated by string theory.

To have a bounce, two conditions need to be met in the Einstein frame: (i) $H_\mathrm{E}(t_{\mathrm{E},0})=0$ at the so-called bounce time $t_{\mathrm{E},0}$, which represents the transition point between contraction and expansion; and (ii) $\dd H_\mathrm{E}/\dd t_\mathrm{E}>0$, meaning the rate of expansion is increasing over time for some interval in $t_\mathrm{E}$, which carries the transition of the Hubble parameter from negative to positive. This precisely corresponds to having the NEC violated for a given period of time.

To get started, let us consider the dilaton to be constant so the dynamics is the same for both frames. Looking at the EOMs when $\dot\phi=0$ [Eqs.~\eqref{eq:OddphiconstFried}--\eqref{eq:OddphiconstPhiEOM}], we notice that a bounce -- at which point $H=0$, which implies that $A(H)$ identically vanishes -- can occur only if $\rho=0$ and $p=-\sigma/(2d)>0$ at that point in time. Considering all these conditions into Eq.~\eqref{eq:NECviolating_matter&corrections}, the NEC-violating condition reduces to $\sigma<0$.
Note, however, that this also means that the `standard' general relativistic NEC in the matter sector has to be violated, i.e., $\rho+p_\mathrm{t}<0$.
Consequently, even when $\alpha'$ corrections are included, a bounce cannot be achieved without violating the matter NEC if $\dot\phi=0$ at that point.

Thus, we need to consider a more general setup where the dilaton is also dynamical if we hope to find non-singular bouncing solutions in the framework of $\alpha'$-complete cosmology. In the Einstein frame, using Eqs.~\eqref{eq:HEFtransf}--\eqref{eq:HEprimegentransf}, one can obtain a bounce where $H_\mathrm{E}=0$ and $\mathrm{d}H_\mathrm{E}/\mathrm{d}t_\mathrm{E}>0$ if
\begin{equation}
    \dot\Phi+H=0\qquad\mathrm{and}\qquad\ddot\Phi+\dot H<0\label{eq:EFbounceconds}
\end{equation}
at that point in time. Substituting the requirement $\dot\Phi=-H$ in the EOMs \eqref{eq:OddFriedwA}--\eqref{eq:OddPhiddwA}, we obtain
\begin{subequations}
\begin{align}
    \ddot\Phi-\frac{d+1}{2}H^2+\frac{A}{2}&=\kappa^2e^{\Phi}\frac{\bar\sigma}{2}\,,\\
    \dot H+H^2-\frac{1}{2d}(\dot HA''+HA')&=\kappa^2e^{\Phi}\bar p\,,\\
    -\frac{d-1}{2}H^2+HA'-\frac{A}{2}&=\kappa^2e^{\Phi}\bar\rho\,.\label{eq:rhocondEFbounce}
\end{align}
\end{subequations}
Summing the above three equations, one can see that the condition $\ddot\Phi+\dot H<0$ becomes equivalent to
\begin{equation}
    \kappa^2 e^{\Phi} (\bar{\rho}+\bar{p}_\mathrm{t})<-(d-1)H^2+\frac{2d-1}{2d}HA'-\frac{1}{2d}\dot HA''\,,\label{eq:EFbouncecond}
\end{equation}
where we recall $\bar p_\mathrm{t}=\bar p+\bar\sigma/2$.
Note that Eq.~\eqref{eq:rhocondEFbounce} can be used to eliminate $\bar{\rho}$. To zeroth order in $\alpha'$ (when $A(H)\equiv 0$), the condition reduces to $p_\mathrm{t}<\rho<0$. This means that, when no $\alpha'$ corrections are included, a bounce point is only achieved if the matter energy density becomes negative at that point (and if the total pressure is even more negative than that). This requires very exotic matter, and the hope is that the $\alpha'$ corrections may allow for the bouncing conditions to be met without the need of exotic matter.

With $\alpha'$ corrections, one may have $\bar\rho>0$ at the bounce point from Eq.~\eqref{eq:rhocondEFbounce} only if
\begin{equation}
    2HA'(H)-A(H)>(d-1)H^2\,.\label{eq:condrhoposbounce}
\end{equation}
It is straightforward to see that if $\alpha'$ corrections are small, then the above condition will be hard to realize.
To see an example where `large' $\alpha'$ corrections might help, consider for instance a function \`a la Dirac-Born-Infeld (DBI)\footnote{The motivation for such an ansatz is discussed at more length in the following subsection.}
\begin{equation}
    A(H)=dH^2\left(1-\sqrt{1-\alpha'H^2}\right)=2dH^2\left(\frac{\alpha'H^2}{4}+\frac{\alpha'^2H^4}{16}+\ldots\right)\,.\label{eq:AHDBI}
\end{equation}
Then, the condition \eqref{eq:condrhoposbounce} for $\bar\rho>0$ reduces to
\begin{equation}
    \frac{26d^2-4d-1-(2d+1)\sqrt{44d^2+4d+1}}{50d^2}<\alpha'H^2<1\,.\label{eq:condrhobarposex}
\end{equation}
The above lower range value is itself in the approximate numerical range $[0.176,0.246]$ for $d\in[3,25]$. 
So typically, as long as $\alpha'H^2\gtrsim 1/4$, one will have $\bar\rho>0$ at the putative bounce point thanks to the $\alpha'$ corrections in the above example.

Working some more with Eq.~\eqref{eq:AHDBI} as an example for $A(H)$, one finds that Eq.~\eqref{eq:EFbouncecond} becomes
\begin{equation}
    w_\mathrm{t}<\frac{2-\mathcal{B}-d\mathcal{B}^4-2\epsilon(1-\mathcal{B})-3(\alpha'H^2)^2(2\epsilon-1)-\alpha'H^2(5-\mathcal{B}-9\epsilon+2\epsilon\mathcal{B})}{\mathcal{B}^2(\mathcal{B}+d(5\alpha'H^2+2\mathcal{B}-3))}\,,
\end{equation}
assuming Eq.~\eqref{eq:condrhobarposex} holds, and where we let $\mathcal{B}:=\sqrt{1-\alpha'H^2}$ to lighten the expression.
Also, we denote the matter EoS by $w_\mathrm{t}=\bar p_\mathrm{t}/\bar\rho$ as before and the string-frame `background EoS' by $\epsilon:=-\dot H/H^2$, akin to the slow-roll parameter in inflation.
The above is a relatively complicated condition, but let us give a few numerical examples. The condition boils down to $w_\mathrm{t}\lesssim-0.5 + \epsilon$ if $\alpha'H^2=1/2$ and $d=3$, $w_\mathrm{t}\lesssim-2.3 + 1.9 \epsilon$ if $\alpha'H^2=1/4$ and $d=3$, and $w_\mathrm{t}\lesssim-44 + 3.8 \epsilon$ if $\alpha'H^2=1/4$ and $d=25$.
In the first couple of examples, a `sensible' background EoS, say $\epsilon\in(1,3)$, yields the condition for the matter EoS to be smaller than a number in the approximate range $[0.5,2.5]$ or $[-0.4,3.5]$, respectively. Thus, standard matter fluids with, e.g., $w_\mathrm{t}\in[0,1]$
could potentially satisfy the Einstein-frame bounce requirement.
However, in the third example (large number of spatial dimensions), the condition is very severe with $w_\mathrm{t}$ having to be very negative, i.e., matter itself strongly violating the NEC.
In summary, this is all case-by-case study, but it looks like there seem to exist sensible situations where standard matter may allow for a NEC-violating, non-singular bounce in the Einstein frame, thanks to the $\alpha'$ corrections.
However, to go beyond the simple observations made here, one needs to study the full dynamics of some chosen ans\"atze, which we turn to in the following subsection.

\paragraph*{A note on string-frame bounces} Beforehand, let us make a short commentary on the possibility of having a bouncing solution in the string frame. The $\mathrm{O}(d,d)$ equations \eqref{eq:OddFriedwA}--\eqref{eq:OddPhiddwA} written in terms of $\phi$ are
\begin{subequations}
\begin{align}
    2\dot\phi^2-2dH\dot\phi+\frac{d(d-1)}{2}H^2+\frac{1}{2}(HA'-A)&=\kappa^2e^{2\phi}\rho\,,\\
    \dot H-2H\dot\phi+dH^2+\frac{1}{2d}\left((2\dot\phi-dH)A'-\dot HA''\right)&=\kappa^2e^{2\phi}p\,,\\
    4\ddot\phi-4\dot\phi^2+4dH\dot\phi-2d\dot H-d(d+1)H^2+A(H)&=\kappa^2e^{2\phi}\sigma\,.
\end{align}
\end{subequations}
In the string frame, if we ask for the conditions for a transition in the scale factor to occur (i.e., a string-frame bounce), then we can check the consistency of the EOMs when $H=0$ and $\dot H>0$,
\begin{subequations}
\begin{align}
    2\dot\phi^2&=\kappa^2e^{2\phi}\rho\,,\label{eq:SFEOMHb3}\\
    \dot H&=\kappa^2e^{2\phi}p\,,\label{eq:SFEOMHb2}\\
    4\ddot\phi-4\dot\phi^2-2d\dot H&=\kappa^2e^{2\phi}\sigma\,,\label{eq:SFEOMHb1}
\end{align}
\end{subequations}
where we recall that $A(H)=A'(H)=A''(H)=0$ when $H=0$.
From the above, it is clear that $\dot H>0$ is only possible if $p>0$. In particular, if $p=0$ (which is the case whenever $p_\mathrm{t}=\sigma/2$ or for vacuum; cf.~Sec.~\ref{sec:linktononOdd}), only the trivial Minkowski solution is obtainable, i.e., $H\equiv 0$ and $\dot H\equiv 0$; a string-frame bounce is certainly not achievable if $p=0$. This is in tension with the results from \cite{Wang:2019kez,Wang:2019dcj}, where non-singular bouncing solutions had been obtained\footnote{The issue with the solutions found in \cite{Wang:2019kez,Wang:2019dcj} appears to be that at the point in time where $H=0$ and $\dot H>0$, one does not recover $F'(H)=0$ as one should. This is due to a problem when inverting the solution for $H(t)$ to find time in terms of $H$: the inversion is ill defined at the point $H=0$, hence the solution cannot continuously hold through $H=0$.} in the absence of any matter sector.

\subsection{Exploring explicit ans\"{a}tze}\label{sec:ansaetze}

In this subsection, we explore the consequences of different $F(H)$ functions. String theory should provide us with the functional form of $F(H)$ if one could compute beta functions to all loops. Since this is currently not reachable, we attempt to guess particular forms for $F(H)$ that respect the symmetries and the lowest-order expansion $F(H)$ should have (cf.~Sec.~\ref{sec:symmetry}), and we see if this ansatz for $F(H)$ yields sensible cosmological background solutions. In the end, we do not attempt to claim that a particular function $F(H)$ is better than another (or that we find the `correct' $F(H)$ function), but we simply aim at exploring the potentiality of the theory. In this approach, a proof of principle for the existence of bouncing solutions supported by the tower of $\alpha'$ corrections might be achieved.

\subsubsection{DBI}

Let us start with the ansatz already introduced in the previous subsection, namely a DBI-like function for $A(H)$ as in \eqref{eq:AHDBI}, or equivalently,
\begin{equation}
    F(H)=-dH^2\sqrt{1-\alpha'H^2}=-dH^2\left(1-\frac{1}{2}\alpha'H^2+\mathcal{O}[(\alpha'H^2)^2]\right)\,.\label{eq:FDBI}
\end{equation}
This function satisfies the lowest-order requirement $F(H)\simeq-dH^2$ and possesses the correct symmetries [invariance under $H\to -H$ and $F(H=0)=0$].
Furthermore, DBI-like terms in the action are ubiquitous in string theory when certain $D$-branes are included (see, e.g., \cite{Leigh:1989jq,Garousi:2000tr,Sen:2003tm,Alishahiha:2004eh}), and the distinctive form of the square root function is peculiar: for it to be real valued, its argument must have a bounded range of validity; consequently, it is notorious for admitting non-singular solutions in string-inspired models and more generally (e.g., \cite{Hinterbichler:2012yn,Fiorini:2013kba,Chamseddine:2016uef}). This naturally begs the question whether the above DBI-like function for $F(H)$, non-perturbative in $\alpha'H^2$, admits non-singular (bouncing) solutions in the context of this paper.

At this point, one can in principle use the above input for $F(H)$, substitute it in the full EOMs and attempt to solve them. In practice, though, this is generally nearly impossible to achieve analytically, and numerically it may be hard to have a global view of the allowed solutions once one chooses arbitrary initial conditions. This is when examining the trajectories in phase space becomes handy. Recalling that the EOMs can be written in the form of \eqref{eq:dynamical_system}, it is clear that the system can reduce to a two-dimensional system of first-order ordinary differential equations of the form
\begin{equation}
    \frac{\dd}{\dd t}\begin{pmatrix}
                        \dot\Phi \\
                        H
                      \end{pmatrix}=\begin{pmatrix}
                                        \mathcal{F}_1(\dot\Phi,H) \\
                                        \mathcal{F}_2(\dot\Phi,H)
                                     \end{pmatrix}\label{eq:2dsys}
\end{equation}
when the RHS of \eqref{eq:dynamical_system} depends only on $\dot\Phi$ and $H$. This can be achieved upon using the constraint equation. In fact, combining Eqs.~\eqref{eq:OddFriedwA}--\eqref{eq:OddPhiddwA} and denoting $w=\bar p/\bar\rho$ and $\lambda=\bar\sigma/\bar\rho$ as before, we can write the EOMs in the above form with
\begin{subequations}\label{eq:F12}
\begin{align}
    \mathcal{F}_1&=\frac{d}{2}\left(1-\frac{\lambda}{2}\right)H^2+\frac{1}{2}\left(1+\frac{\lambda}{2}\right)\left(\dot\Phi^2-A(H)\right)+\frac{\lambda}{4}HA'(H)\,,\\
    \mathcal{F}_2&=\left(1-\frac{A''(H)}{2d}\right)^{-1}\left(\dot\Phi H-\frac{1}{2d}\dot\Phi A'(H)+\frac{w}{2}\left[-dH^2+\dot\Phi^2+HA'(H)-A(H)\right]\right)\,.\label{eq:calF2}
\end{align}
\end{subequations}
Thus, upon specifying the equations of state $w$ and $\lambda$ characterizing the matter sector, all trajectories in phase space can be visualized by looking at the streamflow of the system \eqref{eq:2dsys}.

\begin{figure}
    \centering
    \includegraphics[width=0.48\textwidth]{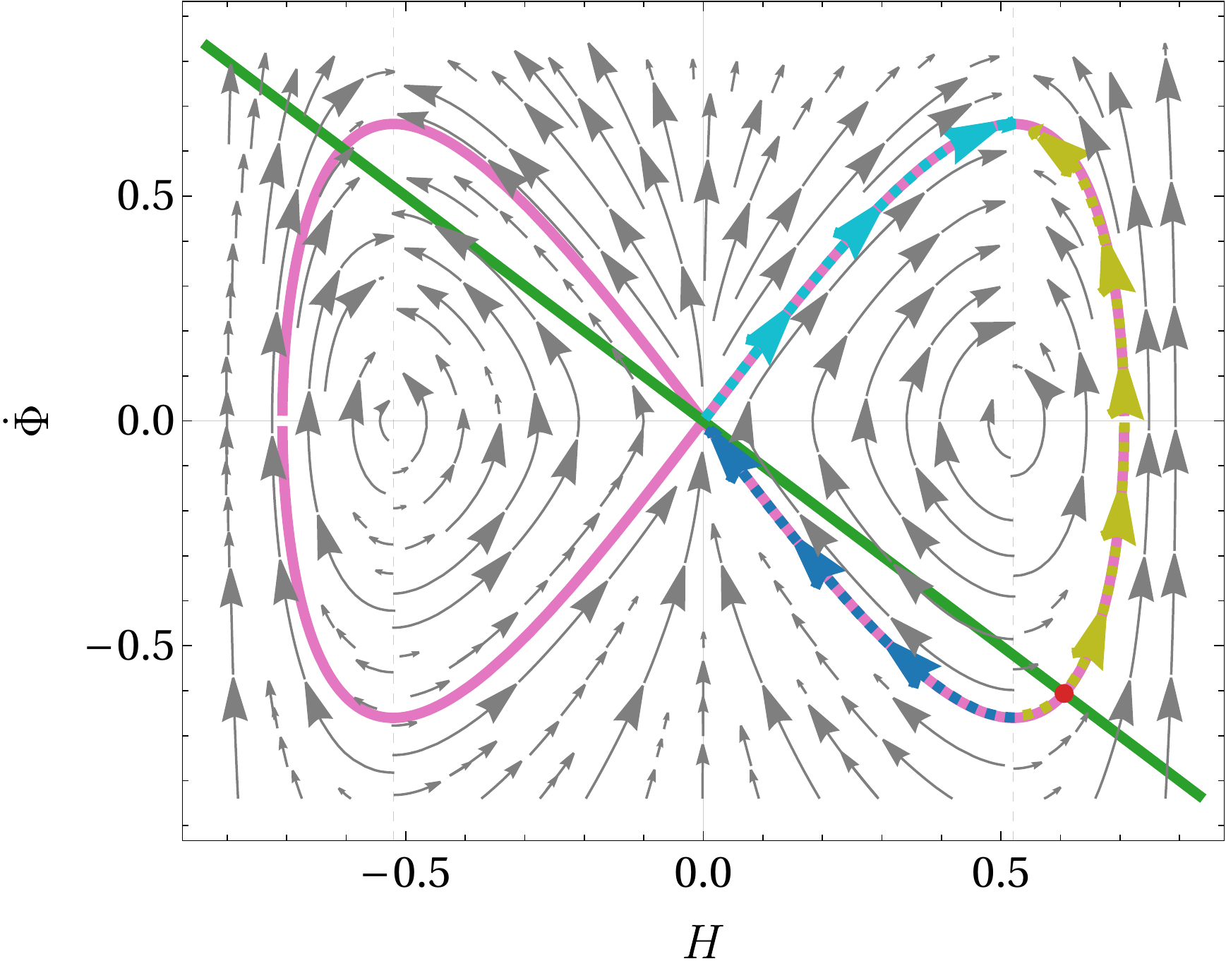}\hspace*{0.1cm}\hfill
    \includegraphics[width=0.48\textwidth]{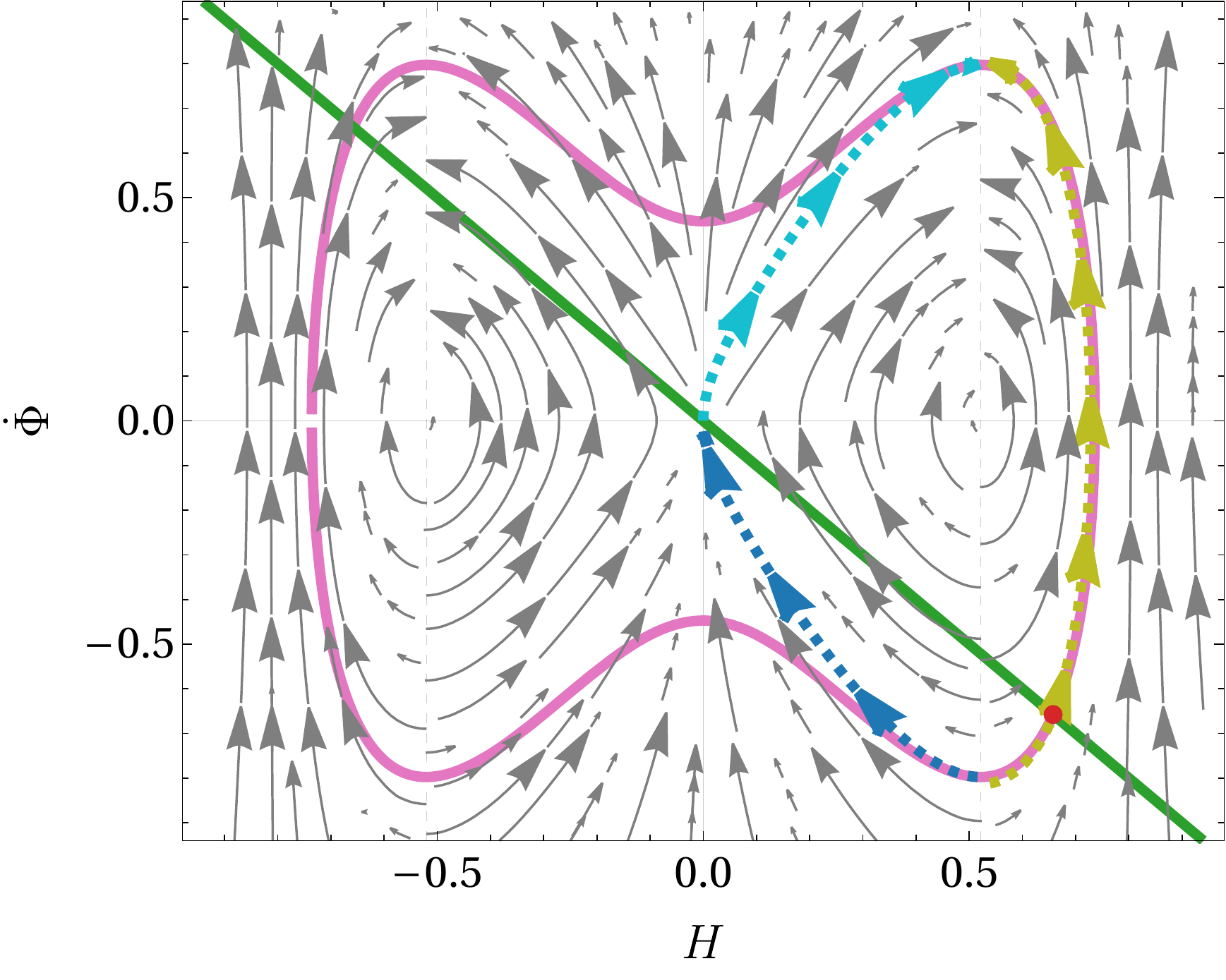}
    \caption{{\small Phase space diagram for the DBI ansatz \eqref{eq:FDBI} obtained from Eqs.~\eqref{eq:2dsys}--\eqref{eq:F12} for matter with $w=\lambda=0$. Units with $\alpha'=1$ are used, and the number of spatial dimensions is set to $d=3$. The pink curve depicts the constraint \eqref{eq:OddFriedwA} in vacuum ($\bar\rho=0$; left plot) and for matter at the `initial' time when $\kappa^2e^{\Phi_0}\bar\rho_0=10^{-1}$ (right plot). The dashed trajectories (in blue, cyan, and olive) show examples of consistent solutions. The green line shows the Einstein-frame bouncing condition $\dot\Phi+H=0$, and the point where the condition is met on a physical trajectory is depicted by a red dot. See text for discussion.}}
    \label{fig:psDBI}
\end{figure}

Taking $A(H)$ in accordance with \eqref{eq:FDBI} and looking at $w=\lambda=0$ as a first example, the trajectories in phase space are shown in Fig.~\ref{fig:psDBI} by the gray arrows. At this point, it is important to mention that the constraint equation is not imposed at the level of the phase space trajectories. Therefore, only a subset of the trajectories actually represent physical solutions to the EOMs. In general, one can search for initial conditions that satisfy the constraint equation \eqref{eq:OddFriedwA}, and those set of valid initial conditions $\{\Phi_0,\dot\Phi_0,H_0\}$ then select specific flow lines in phase space. If we take a vacuum for instance ($\bar\rho\equiv 0$), then the constraint equation reads $\dot\Phi^2-dH^2+HA'(H)-A(H)=0$, and it fully determines the allowed trajectories in $(\dot\Phi,H)$-space. This is depicted by the pink curve in the left plot of Fig.~\ref{fig:psDBI}. As such, if one considers the $H>0$ sector (expansion), then only the blue, cyan, and olive curves are allowed trajectories. Note that only the blue and cyan curves have a well-defined Minkowski limit by reaching the point $(\dot\Phi,H)=(0,0)$, i.e.~the origin. Looking at the blue curve of the bottom right quadrant, we see that the universe starts at a finite value of $H>0$ [some high-energy scale $\sqrt{\alpha'}H\sim\mathcal{O}(1/2)$; we use units with $\alpha'=1$ in the plots] and $\dot\Phi<0$ and expands toward Minkowski with $\dot H<0$. The cyan curve is essentially the reversed process with $\dot H>0$ throughout (it is a `genesis' in the string frame).

If we have non-vacuum matter with\footnote{An example of such pressureless matter in string theory could be a gas of stringy black holes \cite{Quintin:2018loc}.} $w=\lambda=0$ (which implies $w_\mathrm{t}=0$), then we can look at the initial constraint surface (really just a curve here) for some value of $\kappa^2e^\Phi\bar\rho$. In the right plot of Fig.~\ref{fig:psDBI}, the pink curve depicts the example with $\kappa^2e^{\Phi_0}\bar\rho_0=10^{-1}$, where the subscript $0$ means that the function is evaluated at some chosen time $t_0$. In that case, the blue, cyan, and olive curves are just selected examples of allowed trajectories.

At this point, this represents the evolution in the string frame, but one can ask if the corresponding evolution in the Einstein frame admits a non-singular bounce. We recall from \eqref{eq:EFbounceconds} that an Einstein-frame bouncing point when $H_\mathrm{E}=0$ is crossed whenever $\dot\Phi=-H$ in the string frame. This is the meaning of the green curve in all the figures of this section, and we highlight the point where the condition is met on a selected trajectory by a red dot. In the plots of Fig.~\ref{fig:psDBI}, we see that the green line crosses only the olive admissible trajectory. The green line is crossed from the bottom left to the upper right (following the direction of the arrows under forward time evolution), meaning that, at the red dot, one has $\ddot\Phi>0$ and $\dot H>0$, hence $\ddot\Phi+\dot H>0$. This is contrary to the condition of \eqref{eq:EFbounceconds} for this point to represent an Einstein-frame bounce; rather, it seems to be a `turnaround' point, when the universe would go from expansion to contraction. The same happens in the upper left quadrant of the plots: physical trajectories that start on the pink constraint curve cross the green line only with $\ddot\Phi>0$ and $\dot H>0$. Consequently, no Einstein-frame bounce is present in the entire phase space, and it is found that changing the numerical value of $e^{\Phi_0}\bar\rho_0$ in the right plot does not change the conclusion.

\begin{figure}
    \centering
    \includegraphics[width=0.48\textwidth]{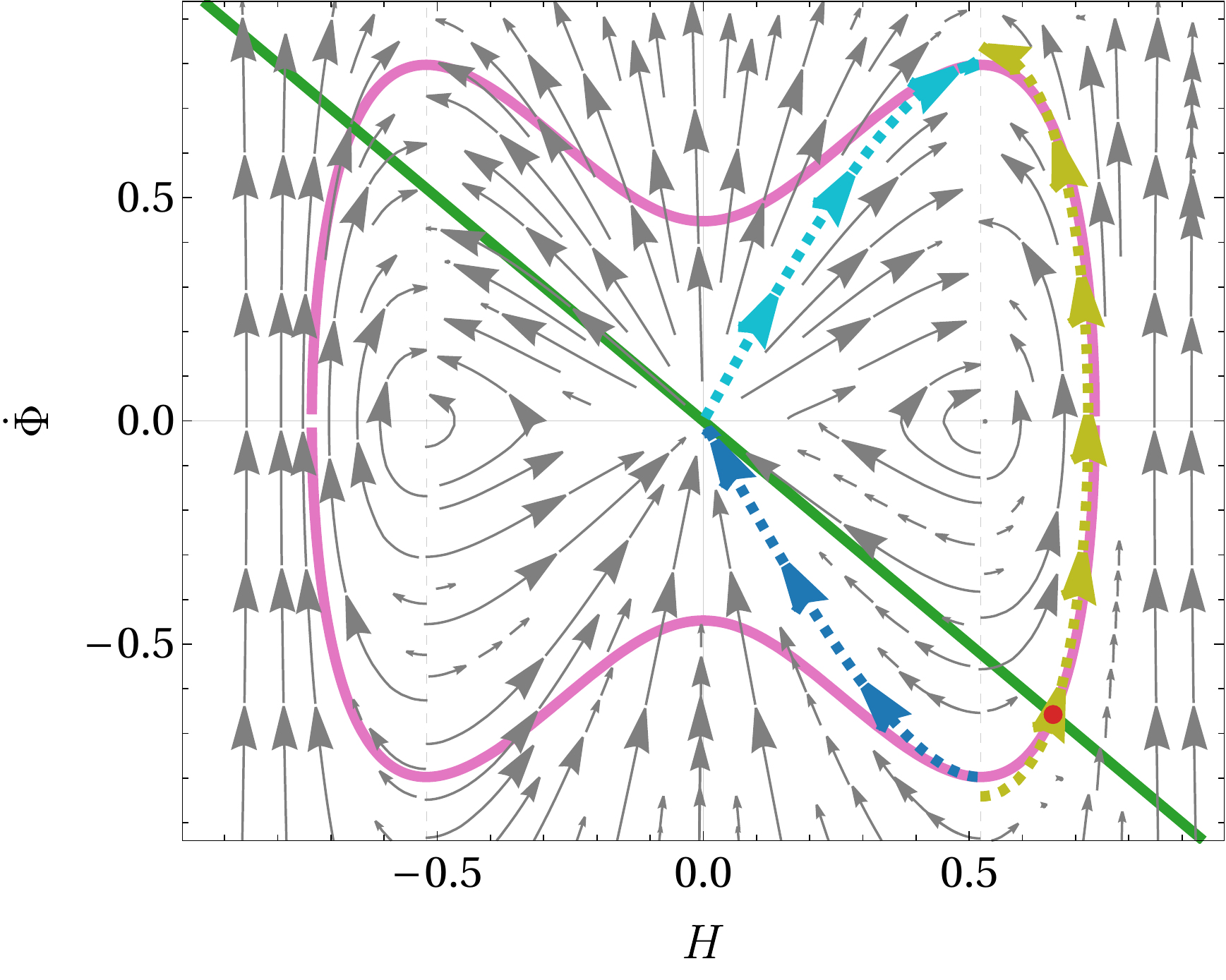}\hspace*{0.1cm}\hfill
    \includegraphics[width=0.48\textwidth]{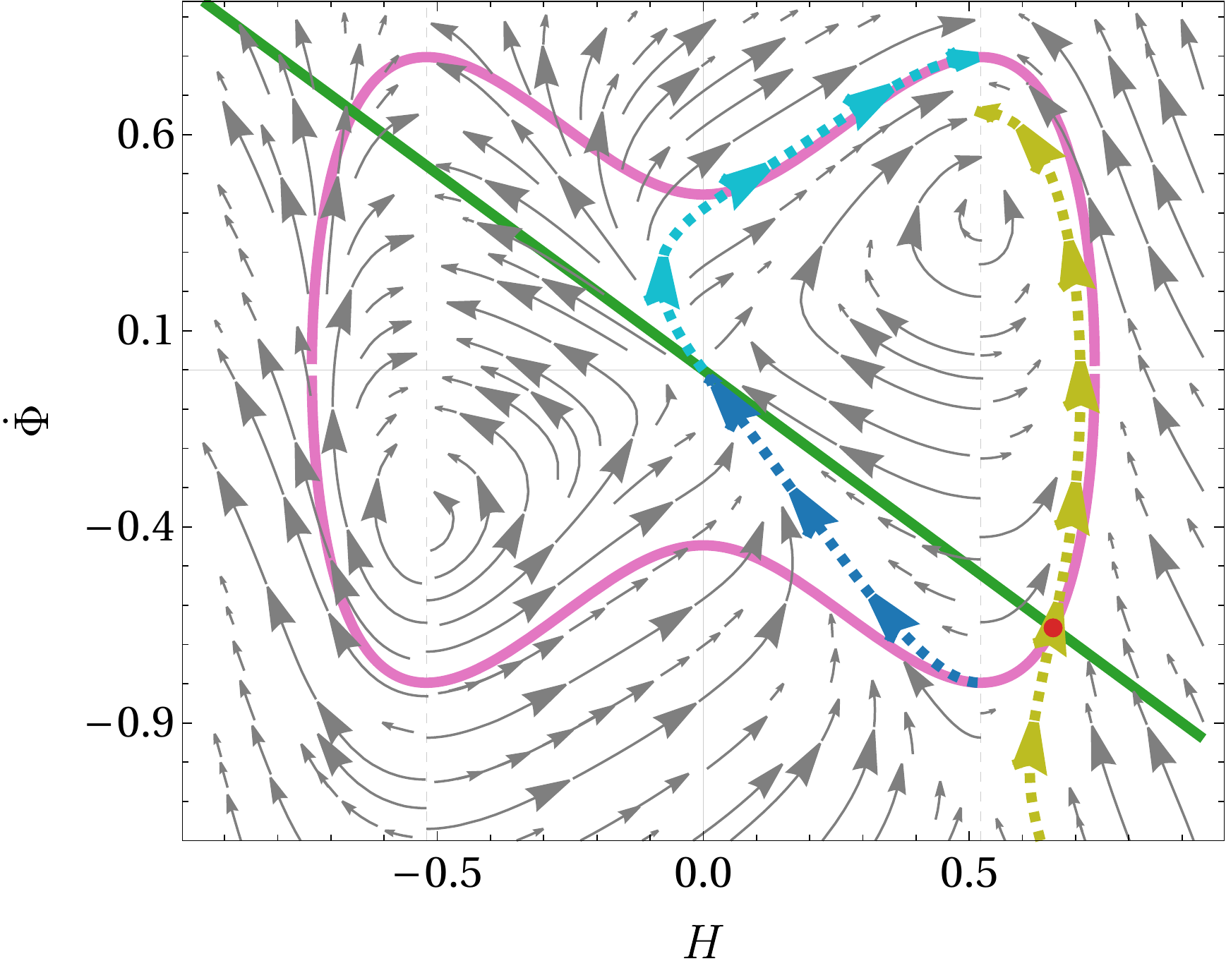}
     \caption{{\small Same plots as Fig.~\ref{fig:psDBI}, except now for $w=0$ and $\lambda=2$ (left plot) and $w=1$ and $\lambda=0$ (right plot).}}
    \label{fig:psDBI2}
\end{figure}

This statement is nevertheless only valid when $w=\lambda=0$, and we can now explore a few other equations of state to see how the phase space is modified. For example, in Fig.~\ref{fig:psDBI2} we show the case where $w=0$ and $\lambda=2$ (left plot) and the case where $w=1$ and $\lambda=0$ (right plot). Note that for both cases $w_\mathrm{t}=w+\lambda/2=1$, which is known as a stiff EoS. The left plot of Fig.~\ref{fig:psDBI2} shows very little differences compared to the right plot of Fig.~\ref{fig:psDBI}, where $\lambda$ was $0$, so it appears coupling to the dilaton does not affect the phase space much in this context. In fact, in all cases, we recover the same general features; in particular, the condition $H_\mathrm{E}=0$ is never satisfied for trajectories that have $\ddot\Phi+\dot H<0$ at that point, meaning a NEC-violating bouncing point in the Einstein frame is not achievable in these cases. A NEC-violating EoS with $w_\mathrm{t}<-1$ could potentially change this conclusion, but this is not the goal here. We rather see that for `reasonable' matter, the tower of $\alpha'$ corrections in the gravitational sector is not sufficient to obtain a non-singular Einstein-frame bounce for a DBI-like $F(H)$ function.

A few more observations are in order. In the right plot of Fig.~\ref{fig:psDBI2}, we see that the phase space is somewhat modified by the appearance of a non-vanishing $\mathrm{O}(d,d)$ pressure ($w\neq 0$). For instance, while the blue trajectory in the right plot is comparable to that in the left plot, the cyan trajectory is somewhat different: it represents a solution that has $H=0$ initially, which is then contracting, bouncing (when it crosses $H=0$ again), and expanding until $\sqrt{\alpha'}H\sim\mathcal{O}(1/2)$. It is thus an example of solution that undergoes a non-singular bounce in the string frame. Nevertheless, the whole solution is not fully free of singularities. Indeed, the vertical dashed gray line around $\sqrt{\alpha'}H\sim\mathcal{O}(1/2)$ (also at its negative value) represents a singularity in the system, and all physical trajectories either start or end there. The mathematical origin of this singularity can be tracked to the point $A''(H)=2d$ in the $\dot H$ EOM [cf.~Eq.~\eqref{eq:calF2} or \eqref{eq:dotHEOM}]. The resulting singularity is somewhat peculiar in that $|H|$ is everywhere bounded, but it is only $\dot H$ (and subsequent higher derivatives) that diverges at that point. As a future singularity, this is known as a sudden future singularity \cite{Barrow:2004xh}; correspondingly, we may call its past equivalent a sudden past singularity, where the universe emerges out of nothing with finite energy, but nevertheless with a divergent initial acceleration. One could also call this a `mild' big bang. Such a singularity (whenever $A''(H)=2d$) appears hard to avoid in the context of this work. That is true even when an actual non-singular bounce is obtained in the Einstein frame, as we will see next.

\subsubsection{Functional renormalization group inspired}

Let us explore another ansatz in this subsection. Specifically, let us choose
\begin{equation}
    A(H)=-1-\frac{23}{12}H^2+\left(\frac{3}{2}+H^2\right)\ln\left(1+\frac{H^2}{2}\right)+\frac{\sqrt{2}}{|H|}\left(1+H^2\right)^{3/2}\mathrm{arctanh}\left(\frac{|H|}{\sqrt{2(1+H^2)}}\right)\,;\label{eq:AFRG}
\end{equation}
accordingly,
\begin{equation}
    F(H)=-dH^2+A(H)=-dH^2\left(1-\frac{29}{80}\alpha'H^2+\mathcal{O}[(\alpha'H^2)^2]\right)\,.
\end{equation}
As before, this function respects the symmetries and the lowest-order requirement.
The inspiration for this specific functional form comes from the work of \cite{Basile:2020xwi,Basile:2021amb} (also explored in \cite{Basile:2021krk}), which uses functional renormalization group techniques to derive $F(H)$.
We remain agnostic about whether or not this fundamentally represents the general functional form of $F(H)$ in string theory, but nevertheless explore some of its consequences from a phenomenological point of view.\footnote{Note that, in fact, the choice \eqref{eq:AFRG} does not exactly represent the function $F(H)$ found in \cite{Basile:2021amb}. Upon removing a cosmological constant and correcting dimensions, our choice of $A(H)$ differs by an overall negative proportionality constant. It is in that sense that \eqref{eq:AFRG} is only phenomenologically inspired by \cite{Basile:2021amb}.}

\begin{figure}
    \centering
    \includegraphics[width=0.48\textwidth]{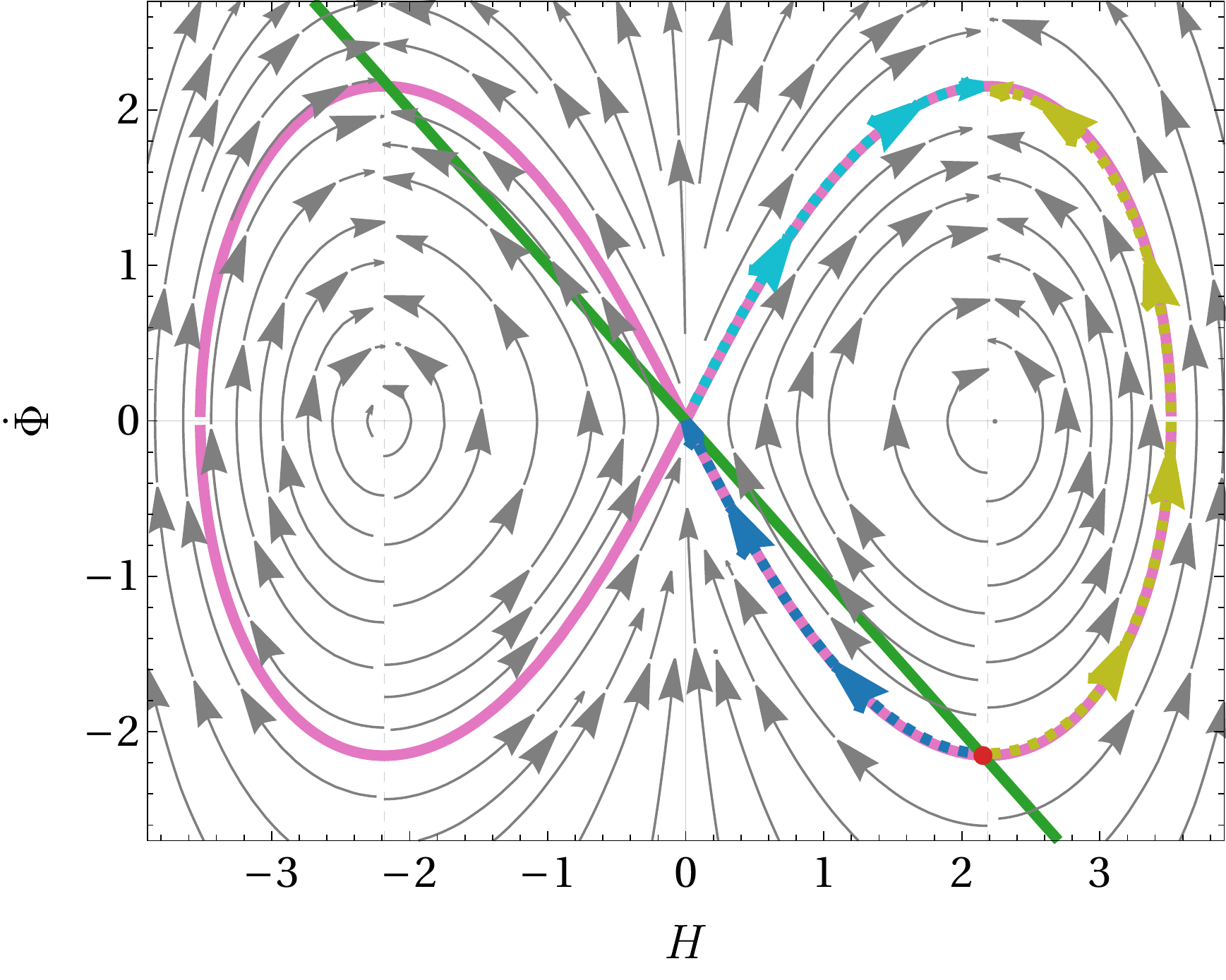}\hspace*{0.1cm}\hfill
    \includegraphics[width=0.48\textwidth]{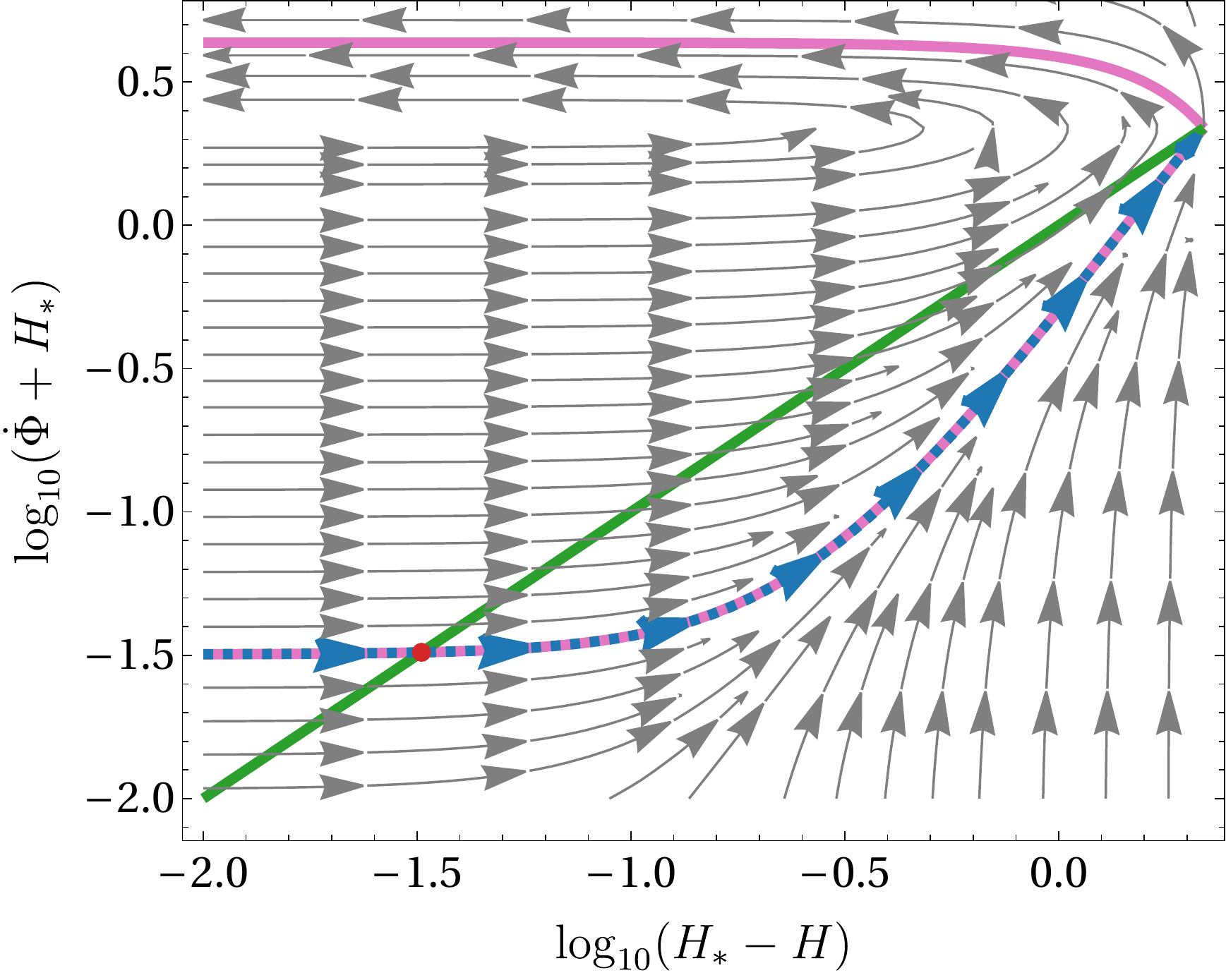}
    \caption{{\small Phase space diagram for the ansatz \eqref{eq:AFRG} obtained from Eqs.~\eqref{eq:2dsys}--\eqref{eq:F12} in vacuum (so with $\bar\rho=w=\lambda=0$). The same conventions as in Fig.~\ref{fig:psDBI} are used. The plot on the right shows the same content as the plot on the left, except that it zooms in closer to the red dot and uses logarithmic scales for both axes. For this illustrative purpose, the horizontal axis in reflected (i.e., it is a function of $-H$), and both axes are shifted by an amount $-H_*$, where $H_*\approx 2.15$ is the point at which the singularity in the EOMs is reached. See text for more details.}}
    \label{fig:psFRGvac}
\end{figure}

Using the same methodology as described earlier, the above function for $A(H)$ allows us to fully describe the system with \eqref{eq:2dsys}--\eqref{eq:F12} and the constraint \eqref{eq:OddFriedwA}. This is first shown in Fig.~\ref{fig:psFRGvac} in the case of vacuum. Interestingly, the overall phase space is not too different from the one obtained with the DBI-like ansatz (Fig.~\ref{fig:psDBI}). Indeed, the overall behavior of the blue, cyan, and olive examples of physical trajectories match one another.
What changes, though, is the location of the point depicted by a red dot where the Einstein-frame bouncing condition $\dot\Phi+H=0$ in green crosses a physical trajectory: it now occurs to the left of the separatrix around $\sqrt{\alpha'}H\approx 2.15$ (the singularity, which we will come back to). This means that a potential Einstein-frame bounce now occurs along a trajectory that has a well-defined Minkowski limit (the blue trajectory reaches the origin where $H=0$), and since the trajectory evolves with $\dot H<0$ (arrows pointing to the left), it may potentially satisfy the Einstein-frame NEC-violating condition $\ddot\Phi+\dot H<0$ at the bounce point. In the left plot of Fig.~\ref{fig:psFRGvac}, the red dot is very close to the separatrix (vertical dashed gray line), which separates, for instance, the blue and olive trajectories. To visualize more clearly that the green line crosses the blue trajectory, we plot on the right of Fig.~\ref{fig:psFRGvac} a zoomed-in version of the left plot. In particular, we use logarithmic scales and move the origin to the point $(H,\dot\Phi)=(H_*,-H_*)$, where $H=H_*\approx 2.15/\sqrt{\alpha'}$ denotes the singular point in the EOMs where $A''(H)=2d$. In this zoomed-in version, it is clear that the red dot falls on the blue trajectory at a point where $\dot H<0$ and where, though $\ddot\Phi>0$, $\ddot\Phi\ll|\dot H|$, hence we can confidently say that $\ddot\Phi+\dot H<0$. Therefore, the conditions of \eqref{eq:EFbounceconds} are satisfied at that point, and so the red dot is a proper bouncing point in the Einstein frame.

\begin{figure}
    \centering
    \includegraphics[width=0.48\textwidth]{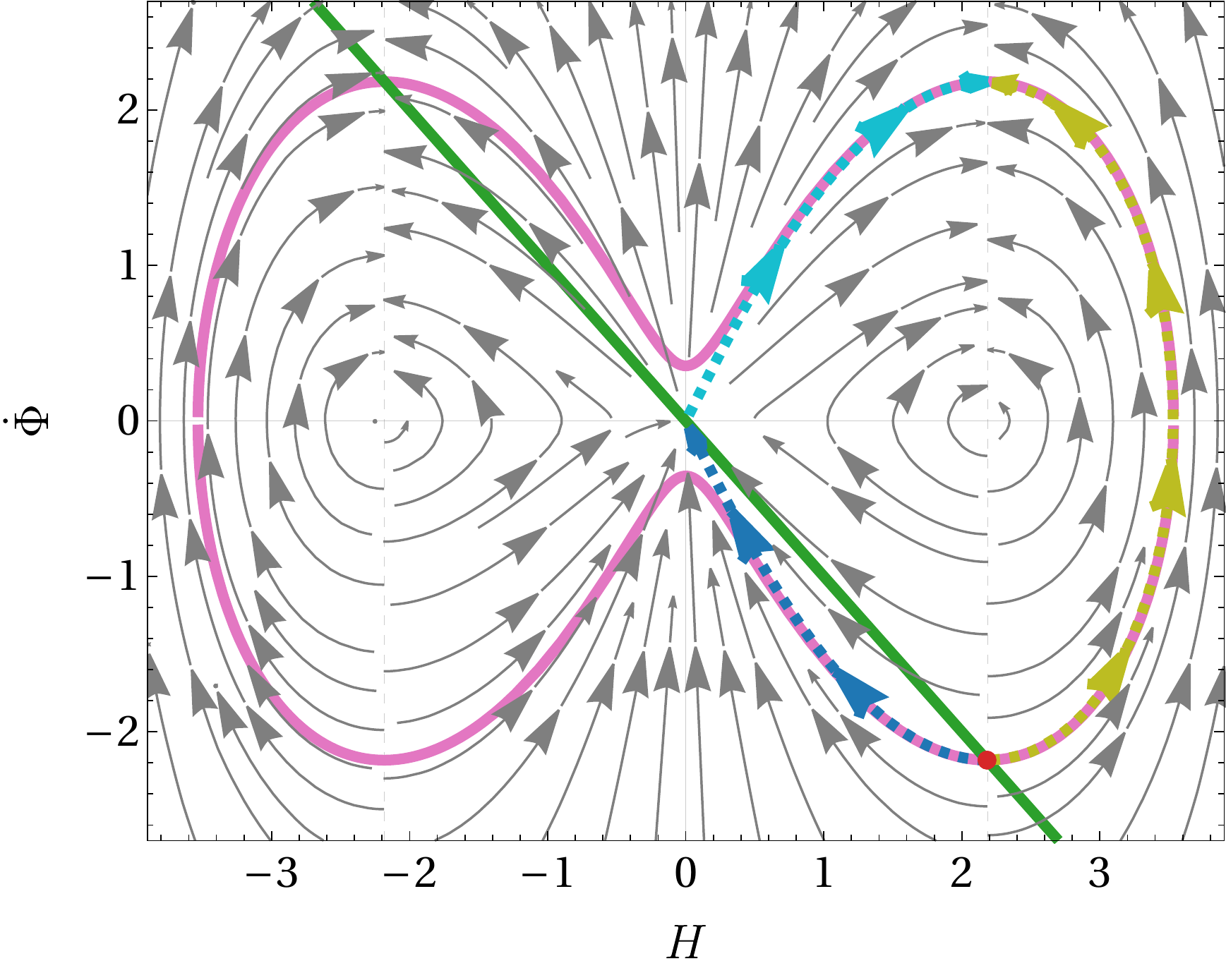}\hspace*{0.1cm}\hfill
    \includegraphics[width=0.48\textwidth]{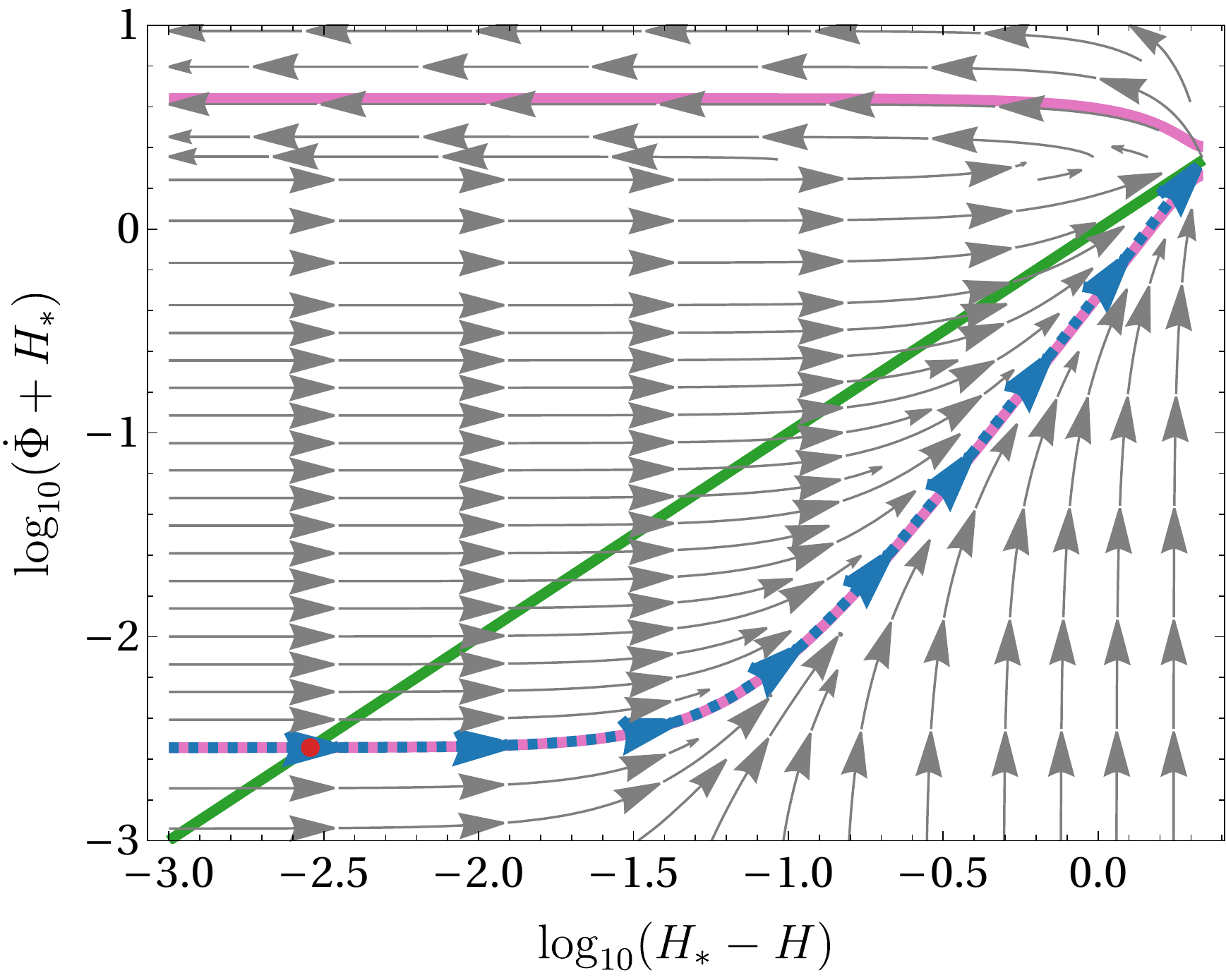}
    \caption{{\small Same plots as Fig.~\ref{fig:psFRGvac}, except now for matter with $w=0$ and $\lambda=2$. The following numerical value is used for the pink `initial' constraint curve: $e^{\Phi_0}\bar\rho_0=10^{-2}$.}}
    \label{fig:psFRGw0l2}
\end{figure}

One can then explore how the addition of matter with various equations of state affects the phase space. Let us show a single such example with $w=0$ and $\lambda=2$ (so $w_\mathrm{t}=1$) in Fig.~\ref{fig:psFRGw0l2}. It is immediately clear upon comparison with Fig.~\ref{fig:psFRGvac} that the changes are very small, qualitatively speaking. In particular, the location of an Einstein-frame bouncing point is again to the left of the separatrix at $\sqrt{\alpha'}H_*\approx 2.15$, along the blue trajectory and such that $\ddot\Phi+\dot H<0$.

\begin{figure}
    \centering
    \includegraphics[width=0.49\textwidth]{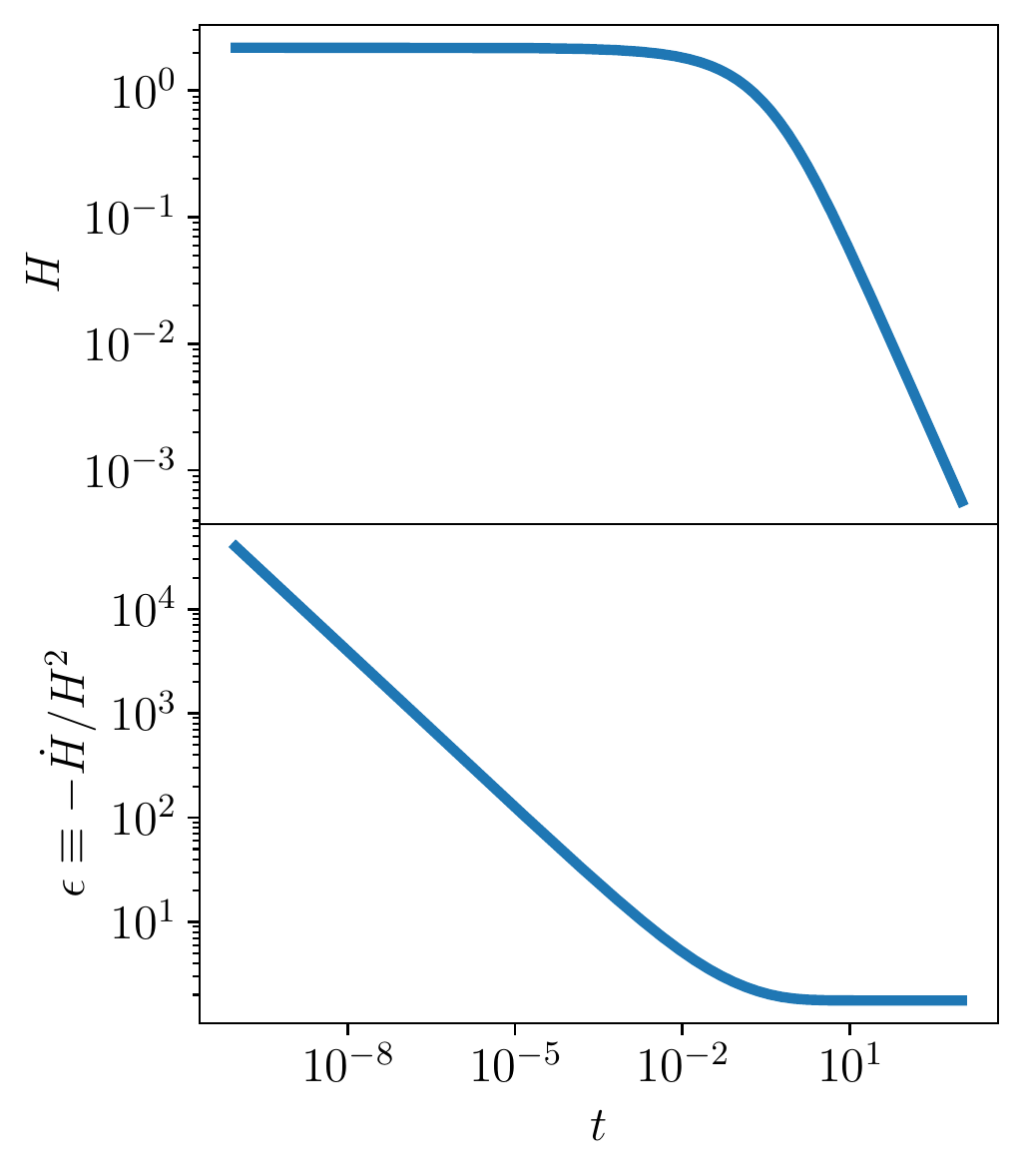}\hfill
    \includegraphics[width=0.49\textwidth]{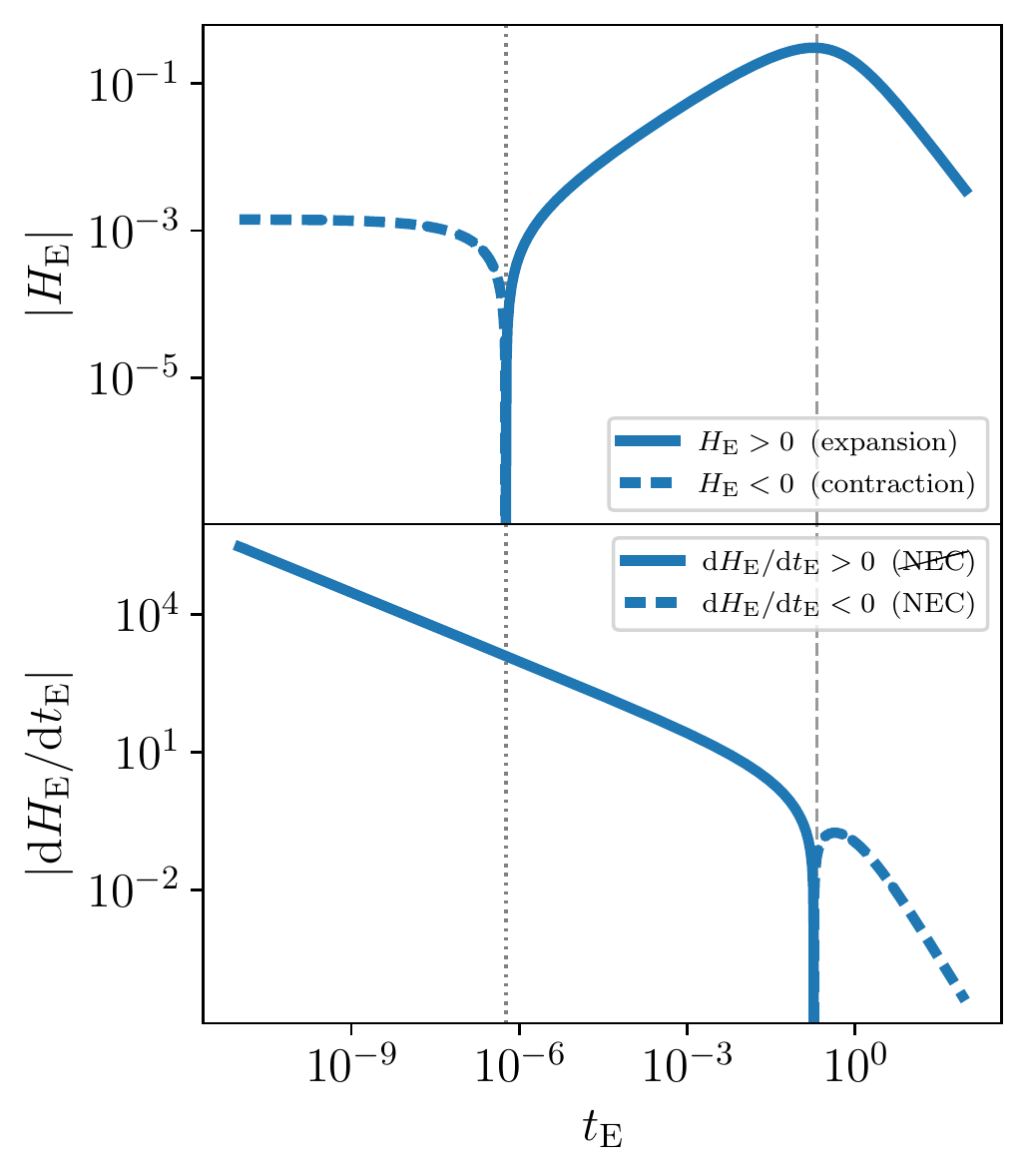}
    \caption{{\small Numerical solution in the string frame (left plots) and in the Einstein frame (right plots) corresponding to the blue dashed trajectory in Fig.~\ref{fig:psFRGw0l2}. The vertical dashed line in the right plots marks the separation between satisfying the NEC and violating it ($\textrm{\cancel{NEC}}$), and the vertical dotted line denotes the bouncing point.}}
    \label{fig:FRGfsol}
\end{figure}

Let us investigate this particular trajectory in more detail. Picking any point along the blue trajectory of Fig.~\ref{fig:psFRGw0l2}, one can numerically solve the full set of EOMs [\eqref{eq:OddwA} with the ansatz \eqref{eq:AFRG} for $A(H)$] forward and backward in time with the specified parameters and units ($d=3$, $w=0$, $\lambda=2$, $e^{\Phi_0}\bar\rho_0=10^{-2}$, $\alpha'=1$). The results are shown in Fig.~\ref{fig:FRGfsol} for the background quantities $H$ and $\epsilon=-\dot H/H^2$ in the string frame and (the absolute value of) $H_\mathrm{E}$ and $\dd H_\mathrm{E}/\dd t_\mathrm{E}$ in the Einstein frame. The conversion from the string frame to the Einstein frame is also done numerically upon integrating \eqref{eq:tEFtransf} and then using Eqs.~\eqref{eq:HEFtransf} and \eqref{eq:HEprimegentransf}.

To the future (as $t,t_\mathrm{E}\to\infty$), we see that $H\to 0^+$, while $\epsilon\to\mathrm{constant}>1$, indicating that one has an expanding FLRW cosmology with an asymptotically constant effective background EoS. This is similarly recovered in the Einstein frame with $H_\mathrm{E}\to 0^+$ and $\dd H_\mathrm{E}/\dd t_\mathrm{E}\to 0^-$, and as expected, the NEC is satisfied in that regime. This is the low-energy regime of the theory. To the past (as $t,t_\mathrm{E}\to 0^+$), we find that $H\nearrow H_*\approx 2.15$, so $H$ is bounded thanks to the infinite tower of $\alpha'$ corrections in the high-curvature regime, but we also find that $\epsilon\to\infty$ (i.e., $\dot H$ diverges to $-\infty$). This is the same kind of divergence as in the previous example with a DBI-like ansatz, namely a sudden past singularity. Once again, the same behavior is found in the Einstein frame, though now with $H_\mathrm{E}$ reaching a negative constant value and $\dd H_\mathrm{E}/\dd t_\mathrm{E}$ diverging to $+\infty$. The reason is that, through the evolution, the universe undergoes a non-singular bounce in the Einstein frame. The bounce point where $H_\mathrm{E}=0$ is depicted by a vertical gray dotted line in the right plot of Fig.~\ref{fig:FRGfsol} around $t_\mathrm{E}\sim 10^{-6}$. In fact, the NEC is violated with $\dd H_\mathrm{E}/\dd t_\mathrm{E}>0$ for the whole regime $t_\mathrm{E}\lesssim 10^{-1}$ (to the left of the vertical gray dashed line). Thus, the model successfully yields a sustained regime of NEC violation such that the Einstein-frame cosmology undergoes a non-singular bounce, but the whole cosmology remains singular due to the sudden past singularity. A more successful model would need to exit the NEC-violating phase to the past in a similar fashion to what it does to the future. However, the present ansatz does not allow for such a fully non-singular background, and it remains unknown whether any $F(H)$ could allow for a fully singularity-free solution in that manner. Practically, trajectories in phase space would need to `bend' such that a point where $A''(H)=2d$ is never reached. How this may be achieved remains unknown at this point.

\section{Discussion and conclusions}\label{sec:discussion}

In the current paper, we first revisited some recent developments in $\alpha'$-complete string cosmology. This allowed us to clarify the relation between the $\mathrm{O}(d,d)$ pressure (that naturally arises from an $\mathrm{O}(d,d)$-covariant approach) and the total pressure that is more common in dilaton gravity. We also clarified how Einstein gravity may be recovered at low curvature, i.e., what matter content allows for late-time Friedmann equations with fixed or decaying dilaton. Moving on to solutions that encapsulate the whole tower of $\alpha'$ corrections, we showed that the linear dilaton CFT is recovered, while for the case of a fixed dilaton, we showed that the allowed solution space is very constrained when the $\mathrm{O}(d,d)$ pressure vanishes: only Minkowski and dS solutions are allowed. In vacuum, recovering Minkowski is expected, but the possibility of dS solutions is remarkable and deserves further discussion.

From the results of Sec.~\ref{sec:constantdilaton}, the duality-invariant framework allows for vacuum dS solutions in the Einstein frame, provided the function $F(H)$ satisfies some conditions, which also restrict the possible values of the dS radius. Given the present ignorance about fully non-perturbative time-dependent backgrounds in string theory, it is currently unknown whether string theory would supply an $F(H)$ that satisfies such conditions. Recently, the existence of dS backgrounds in string theory has been intensively debated (e.g., \cite{Moritz:2017xto,Danielsson:2018ztv,Dasgupta:2018rtp,Gautason:2018gln,Hamada:2018qef,Kallosh:2019axr,Hamada:2019ack,Gautason:2019jwq,Kachru:2019dvo}). The swampland program (see \cite{Palti:2019pca} for a review) aims to identify low-energy effective field theories that cannot be consistently coupled with string theory. In particular, the dS conjecture states that the potential for the scalar fields parametrizing a given string compactification construction is not flat enough to sustain a meta-stable dS background \cite{Obied:2018sgi,Garg:2018reu, Ooguri:2018wrx}. However, these conjectures are not in direct conflict with an admissible function $F(H)$ that would satisfy the conditions for the existence of dS: as discussed in Sec.~\ref{sec:constantdilaton}, the non-perturbative dS solutions from $\alpha'$-complete cosmology are likely to have a dS radius of the order of the string length and so are not in tension with the swampland program that constrains low-energy solutions with energy scales much smaller than the string scale.

The non-perturbative dS solutions here require the whole tower of $\alpha'$ corrections to exist, i.e., they cannot ever be described by any truncation of the fully corrected theory. In that sense, there is no `effective description' of such backgrounds, as they are not solutions to the low-energy effective field theory (supergravity). A very similar conclusion is also found in the M-theory analysis of \cite{Dasgupta:2019gcd,Dasgupta:2019vjn,Brahma:2020htg,Brahma:2020tak,Bernardo:2020lar}, in which the authors claim that the only way to have a dS background with a hierarchy between the terms in the tower of higher-order corrections (in the curvature and fluxes) is for the fluxes and internal cycles to be time dependent. Moreover, if all the fluxes are made time independent or even turned off, then the only way to get a dS solution in their approach is for all the corrections to contribute with the same scaling, meaning a breakdown of the effective description, resonating with what we obtain from the duality invariant framework. In light of these remarks, it is possible that the swampland conjectures are not in conflict with the existence of non-perturbative dS backgrounds in string theory.

The second part of this paper focused on the exploration of the violation of typical relativistic energy conditions and the corresponding consequences. This allowed us to show that for solutions with constant $H$ of the form \eqref{eq:H0varPhisol}, when the $\mathrm{O}(d,d)$ pressure vanishes and excluding the possibility of including a cosmological constant, violation of the effective SEC is impossible, meaning that accelerated expansion in the Einstein frame can never be achieved. If an $\mathrm{O}(d,d)$ pressure is included, then the situation changes drastically: not only are accelerating Einstein-frame solutions admissible, they can moreover be achieved with positive total pressure thanks to the tower of $\alpha'$ corrections. It would thus be interesting to see if one could find a sensible stringy matter Lagrangian, which has a non-vanishing $\mathrm{O}(d,d)$ pressure and positive total pressure, that would allow for Einstein-frame accelerated expansion (in the very early universe for instance).

We similarly derived the conditions for violating the effective NEC in the Einstein frame. While the conditions are generally quite complicated, we estimated that when one enters the high-curvature regime with $\alpha'H^2\sim\mathcal{O}(1)$, reasonable matter with positive energy density and pressure could in principle allow for NEC violation, once again thanks to the tower of $\alpha'$ corrections. We then explored a couple of phenomenologically motivated non-perturbative $F(H)$ functions that allowed us to describe the whole phase space of permitted solutions. In doing so, we found interesting solutions such as non-singular bouncing cosmologies, either in the string frame or in the Einstein frame. However, such solutions are far from being generic; quite the contrary, they rather appear hard to find, and moreover, we could not find solutions that remained non-singular throughout time. Indeed, most solutions remained plagued with sudden -- though mild -- singularities. Nevertheless, the non-singular bouncing toy models that were found serve as a proof of principle that an infinite tower of $\alpha'$ corrections may allow for singularity resolution in cosmological backgrounds.

These conclusions are restricted to our exploration of a few $F(H)$ functions. Of course, this is far from being exhaustive, and so one cannot reach generic conclusions. There may well exist a particular functional form for $F(H)$ that would yield fully singularity-free cosmologies. However, our small exploration showed that `simple' $F(H)$ functions could not completely resolve singularities, which might suggest that $\alpha'$ corrections (even to all orders) are not sufficient to properly describe the physics all the way to very high-energy scales. For instance, our work certainly did not include any loop corrections, i.e., we always remained in the weak-coupling regime. Such corrections are certainly expected to introduce yet more higher-curvature corrections (see, e.g., \cite{Gasperini:2002bn,Gasperini:2007zz} are references therein), but those would effectively just change the functional form of $F(H)$, so loop corrections are perhaps not sufficient to fully resolve singularities. Our work also did not explore the possibility of a potential for the $\mathrm{O}(d,d)$ dilaton, $V(\Phi)$. While it is not clear if this is fundamentally allowed in string theory, it certainly respects $\mathrm{O}(d,d)$ covariance -- with regard to spacetime covariance, this is said to be a non-local potential (see \cite{Meissner:1991zj,Gasperini:1991ak,Gasperini:1996np,Gasperini:2002bn,Gasperini:2007ar,Gasperini:2007zz}). At zeroth order in $\alpha'$, it is already known that the appropriate choice of potential allows for singularity resolution (see \cite{Gasperini:1996np,Gasperini:2002bn,Gasperini:2007ar,Gasperini:2007zz}), so it would be interesting to see how this generalizes with the addition of an infinite tower of $\alpha'$ corrections. Whether or not a non-local dilaton potential is the way to go, non-locality is certainly an interesting feature of string theory that is hard to capture at the effective field theory level, and which may well be key to properly resolve singularities in string theory. Some other approaches in that direction worthy of mention include infinite-derivative gravity (see, e.g., \cite{Biswas:2005qr,Biswas:2010zk} and subsequent works) and S-branes (e.g., \cite{Kounnas:2011fk,Florakis:2010is,Brandenberger:2013zea,Brandenberger:2020eyf}).

Let us end by commenting on the applicability of $\alpha'$-complete string cosmology to actual scenarios of the very early universe. Recent developments, including this current work, certainly pave the way to finding new stringy descriptions of the early universe. Beyond inflation, it may lead to developments in string gas cosmology (see \cite{Bernardo:2020nol,Bernardo:2020bpa}), but it would also be interesting to apply the framework to other alternative approaches, including pre-big bang cosmology (as envisioned in \cite{Gasperini:1996fu}) and complementary stringy descriptions of the high-density state of the very early universe such as in the stringy black hole gas scenario \cite{Quintin:2018loc}. To go beyond background models, though, lies an important challenge. Indeed, the assumptions of homogeneity and isotropy are paramount to the whole program of stringy $\alpha'$ corrections to all orders in cosmology, and a proper description of the very early universe has to be able to tackle cosmological perturbations. In particular, going beyond homogeneity and including spatial dependence shall be a major obstacle to circumvent for the $\alpha'$-complete program to become phenomenologically reliable after including perturbations.

\vskip23pt
\subsection*{Acknowledgments}
We thank Jean-Luc Lehners for his involvement in the early stages of this work and for valuable discussions and insightful comments that shaped many aspects of this work.
J.\,Q.~also thanks Olaf Hohm and Jeong-Hyuck Park for enlightening discussions during the Geometry and Duality Workshop held at the Albert Einstein Institute (AEI) in December 2019.
G.\,F.~thanks the AEI for hospitality
and J.\,Q.~and H.\,B.~thank the stimulating atmosphere at Nordita during the beginning of this project.
Research at the AEI is supported by the European Research Council (ERC) in the form of the ERC Consolidator Grant CoG 772295 `Qosmology'.
J.\,Q.~further acknowledges financial support in part from the \textit{Fond de recherche du Qu\'ebec --- Nature et technologies} postdoctoral research scholarship and the Natural Sciences and Engineering Research Council (NSERC) of Canada Postdoctoral Fellowship.
H.\,B.'s research is also partially supported by funds from NSERC.

\clearpage 
\appendix

\section{The general transformation to the Einstein frame}\label{app:EF}

The general idea of the Einstein frame (EF) is to take the stringy action
\begin{equation}
    S^{(\mathrm{SF})}[\bm{G},\phi,\cdots]=\frac{1}{2\kappa^2}\int\dd^Dx\,\sqrt{-G}e^{-2\phi}\left(R[\bm{G}]+4G^{\mu\nu}\partial_\mu\phi\partial_\nu\phi+\cdots\right)\,,
\end{equation}
which is said to define the string frame (SF), and transform it via the Weyl field redefinition (the sub/superscript `E' indicates an Einstein-frame quantity; it is in the string frame otherwise)
\begin{equation}\label{Weylredefinition}
    G_{\mu\nu}^{\mathrm{E}}(x)=e^{-\frac{4\phi(x)}{D-2}}G_{\mu\nu}(x)\,,
\end{equation}
such that the second derivative part of the action for $G_{\mu\nu}^{\mathrm{E}}$ is the Einstein-Hilbert action, 
\begin{equation}
    S^{(\mathrm{EF})}[\bm{G}_\mathrm{E},\phi_\mathrm{E},\cdots]=\int\dd^Dx\,\sqrt{-G_\mathrm{E}}\left(\frac{M_\mathrm{Pl}^{D-2}}{2}R[\bm{G}_\mathrm{E}]-\frac{1}{2}G_\mathrm{E}^{\mu\nu}\partial_\mu\phi_\mathrm{E}\partial_\nu\phi_\mathrm{E}+\cdots\right)\,,\label{eq:EFaction0thorder}
\end{equation}
where the Einstein-frame dilaton $\phi_\mathrm{E}$ is now a dimensionful, minimally coupled scalar field with canonical kinetic term.

Starting with a flat FLRW metric in the string frame, Eq.~\eqref{Weylredefinition} implies that the Einstein-frame metric is also of the FLRW form, with lapse function and scale factors given by
\begin{equation}
    n_{\mathrm{E}}(t_\mathrm{E})\,\dd t_\mathrm{E} = e^{-\frac{2\phi(t)}{D-2}} n(t)\,\dd t\,, \qquad a_{\mathrm{E}}(t_\mathrm{E}) = e^{-\frac{2\phi(t)}{D-2}}a(t)\,.
\end{equation}
Choosing the lapse functions in both frames to be unity, the Einstein-frame time $t_\mathrm{E}$ is fixed by
\begin{equation}\label{eq:tEFtransf}
    t_{\mathrm{E}}(t) - t_{\mathrm{E},0} = \int_{t_0}^{t} \dd \tilde{t}\, e^{-\frac{2\phi(\tilde{t})}{D-2}}\,,
\end{equation}
where $t_{\mathrm{E}, 0} := t_{\mathrm{E}}(t_0)$. For a given solution in the string frame, one may solve this integral to find $t(t_{\mathrm{E}})$ and write $a_{\mathrm{E}}(t_{\mathrm{E}})$. The Hubble rate in the Einstein frame is related to the string-frame variables as
\begin{equation}
    H_{\mathrm{E}} := \frac{\dd\ln a_{\mathrm{E}}}{\dd t_\mathrm{E}}= e^{\frac{2\phi}{D-2}}\frac{\dd\ln a_\mathrm{E}}{\dd t} = e^{\frac{2\phi}{D-2}} \left(H - \frac{2\dot{\phi}}{D-2}\right)\,,
\end{equation}
and recalling $e^{-\Phi}=a^de^{-2\phi}$, this can be written as
\begin{equation}\label{eq:HEFtransf}
    H_{\mathrm{E}} = -\frac{e^{\frac{2\phi}{d-1}}}{d-1}\left(H +\dot{\Phi}\right) = -\frac{a^{\frac{d}{d-1}}e^{\frac{\Phi}{d-1}}}{d-1}\left(H+\dot{\Phi}\right)\,.
\end{equation}
Similar algebra yields
\begin{subequations}
\begin{align}
    &\frac{\dd^2a_\mathrm{E}}{\dd t_\mathrm{E}^2}=ae^{\frac{2\phi}{d-1}}\left(\frac{\ddot a}{a}-\frac{2}{d-1}\frac{\dot a}{a}\dot\phi-\frac{2}{d-1}\ddot\phi\right)=-\frac{a^{\frac{2d-1}{d-1}}e^{\frac{\Phi}{d-1}}}{d-1}\left(\dot H+\ddot\Phi+H(H+\dot\Phi)\right)\,,\label{eq:appEFtranf}\\
    &\frac{\dd H_\mathrm{E}}{\dd t_\mathrm{E}}=-\frac{a^{\frac{2d}{d-1}}e^{\frac{2\Phi}{d-1}}}{d-1}\left[\dot H+\ddot\Phi+(H+\dot\Phi)\left(H+\frac{H+\dot\Phi}{d-1}\right)\right]\,,\label{eq:HEprimegentransf}
\end{align}
\end{subequations}
which are useful formulae throughout this work.

\newpage 
\phantomsection
\addcontentsline{toc}{section}{References}

\let\oldbibliography\thebibliography
\renewcommand{\thebibliography}[1]{
  \oldbibliography{#1}
  \setlength{\parskip}{0pt}
  \setlength{\itemsep}{0pt} 
  \footnotesize 
}

\bibliographystyle{JHEP2}
\bibliography{alphaprimerefs}

\end{document}